\documentclass[twocolumn]{aastex631}
\usepackage{bm}

\newcommand{\mytilde}{\raise.19ex\hbox{$\scriptstyle\sim$}}

\usepackage[utf8]{inputenc}
\usepackage{CJK}
\usepackage{amsmath}
\usepackage{tabularx}
\usepackage{hyperref}
\usepackage{newunicodechar}
\newunicodechar{Λ}{\Lambda}
\usepackage{multirow}
\usepackage{xcolor}
\newunicodechar{−}{\ensuremath{-}}
\usepackage{soul}

\shorttitle{Bullet Cluster Is a Minor Merger}

\begin{document}

\title{Joint JWST--DECam Lensing Reveals That the Bullet Cluster Is a Minor Merger}

\correspondingauthor{M. James Jee}
\email{boseongcho@yonsei.ac.kr, mkjee@yonsei.ac.kr}

\author[0009-0009-4273-6132]{Boseong Young Cho}
\affiliation{Department of Astronomy, Yonsei University, 50 Yonsei-ro, Seoul 03722, Korea}
\author[0000-0002-5751-3697]{M. James Jee}
\affiliation{Department of Astronomy, Yonsei University, 50 Yonsei-ro, Seoul 03722, Korea}
\affiliation{Department of Physics and Astronomy, University of California, Davis, One Shields Avenue, Davis, CA 95616, USA}
\author[0000-0001-9139-5455]{Hyungjin Joo}
\affiliation{Department of Astronomy, Yonsei University, 50 Yonsei-ro, Seoul 03722, Korea}
\author[0000-0001-7148-6915]{Sangjun Cha}
\affiliation{Department of Astronomy, Yonsei University, 50 Yonsei-ro, Seoul 03722, Korea}
\author[0000-0002-2550-5545]{Kim HyeongHan}
\affiliation{Department of Physics, Duke University, Durham, NC 27708, USA}
\affiliation{Department of Astronomy, Yonsei University, 50 Yonsei-ro, Seoul 03722, Korea}

\begin{abstract}
We present the first robust virial masses of the Bullet Cluster's three individual components
from a joint weak+strong lensing analysis combining JWST/NIRCam and DECam observations. Despite its status as the benchmark system for dark matter and merger studies, inferred mass ratios for the Bullet Cluster have spanned a wide range from $\mytilde$2:1 to
$\gtrsim$10:1 over more than two decades. We revisit this tension through three key advances: (1) JWST’s exceptional data quality enables us to resolve three distinct halos, (2) DECam’s wide-field coverage beyond its virial radius eliminates the need for extrapolation, and (3) high-fidelity strong-lensing priors mitigate weak-lensing model bias. We obtain $M_{200c} = 15.11^{+2.48}_{-2.10} \times 10^{14}M_{\odot}$ for the main cluster and $1.49^{+0.32}_{-0.25} \times 10^{14}M_{\odot}$ for the subcluster, yielding a mass ratio of $10.14^{+3.22}_{-2.47}$, definitively classifying the Bullet Cluster as a minor merger.
This result reconciles the long-standing tension in the mass ratio and provides updated initial parameters for future modeling of this iconic system.
\end{abstract}

\section{Introduction}
\label{sec:int}

Galaxy cluster mergers are the most energetic events in the universe since the Big Bang, releasing up to $\mytilde10^{64}$ ergs \citep[e.g.,][]{Ricker_2001} with collision velocities
of the order of a few thousand km~s$^{-1}$ 
\citep[e.g.,][]{Sarazin_2002, Markevitch_2004}. These extreme astrophysical laboratories offer unique insights into dark matter, plasma dynamics, and structure formation that cannot be replicated on Earth \citep[e.g.,][]{Harvey_2015, Kim_2017, Jee_2026}.

The Bullet Cluster (1E 0657-56) at $z=0.296$ exemplifies these phenomena most
strikingly, displaying the clearest observed mass-gas dissociation and providing direct evidence for dark matter's existence \citep[e.g.,][]{Clowe_2004, Clowe_2006, Markevitch_2004}. Its unique properties, nearly
plane-of-sky orientation (viewing angle $<10^{\circ}$), extreme collision velocity ($\mytilde4700$~km~s$^{-1}$), and exceptional bow shock features, have made it the primary testing ground for dark matter self-interaction cross-sections, serving as the benchmark for hydrodynamical simulations of cluster mergers \citep[e.g.,][]{Markevitch_2004, Springel_2007, Randall_2008, Robertson_2017}.

Despite more than two decades of intensive study, a fundamental challenge persists: the system's virial masses remain poorly constrained.
Observed mass ratios vary widely from $\mytilde$2:1 \citep{Paraficz_2016, Richard_2021} to $\mytilde$100:1 \citep{Barrena_2002}. 
This wide range highlights the persistent tension between simulation-based expectations and observational inferences.
Hydrodynamical simulations that successfully reproduce the observed mass distribution and gas dynamics, including the bow shock,
mostly require minor-merger initial conditions with mass ratios
of order 10:1
\citep[e.g.,][]{Milosavljevic_2007, Springel_2007, Mastropietro_2008, Lage_2014}. In contrast, observational studies, particularly those employing strong lensing (SL) analyses that do not rely on equilibrium assumptions, consistently imply that the Bullet Cluster is a major-merger system with
a mass ratio of $\mytilde$2:1.

This tension fundamentally limits the utility of this iconic merger system for dark matter studies based on numerical simulations. For instance, \citet{Randall_2008} constrained the dark matter self-interaction cross-section through numerical modeling of the Bullet Cluster, adopting initial halo masses varying from $\mytilde$1.4:1 to $\mytilde$1.9:1\footnote{Estimated using their initial conditions listed in Table 1.} mainly motivated by the SL results of \cite{Bradac_2006}.
Because such simulations depend sensitively on these assumed initial conditions, the persistent uncertainty in the true mass ratio has limited their constraining power. 

The uncertainty
arises from methodological limitations inherent to each observational approach. 
Techniques
assuming dynamical or hydrostatic equilibrium
become unreliable
for such a
system undergoing a violent merger: \citet{Barrena_2002} obtained an
extremely high mass ratio of $\mytilde$100:1
based on velocity dispersions. 
In contrast, gravitational lensing determines the projected mass distribution without invoking dynamical or hydrostatic equilibrium, although both strong- and weak-lensing rely on modeling assumptions to infer a spherically enclosed mass, and these assumptions introduce systematic uncertainties when either technique is used in isolation.

SL analyses provide high-accuracy and high-precision mass constraints at small radii but rapidly lose constraining power beyond the Einstein radius, requiring substantial extrapolation to estimate the virial mass.
Previous weak-lensing (WL) studies using ground-based data with source densities of $\mytilde5$--$15$~arcmin$^{-2}$ were unable to constrain the subcluster's virial mass, obtaining either projected masses for the subcluster \citep{Clowe_2004} or only the total cluster virial mass through Navarro-Frenk-White (NFW; \citealt{Navarro_1996}) profile fitting \citep{Melchior_2015}. Although \citet{Clowe_2006} conducted a WL analysis using Hubble Space Telescope (HST) data, they did not determine virial masses. Furthermore, WL analysis of merging clusters inherently suffers from model bias because it typically assumes
an analytic profile, such as the NFW. \citet{Lee_2023} demonstrate that, unless a strong prior on concentration is available,
fitting NFW profiles to merging clusters can introduce biases up to $\mytilde$60\%, as the merger significantly alters the halo properties from average expectations.

Also, accurately identifying the number of distinct halos is critical for reliable mass estimation of merging clusters. Misidentifying substructures, for instance, modeling only one halo when two truly exist, can introduce substantial model bias and lead to misinterpretation of shear signals. Such biases propagate into derived quantities such as virial masses, concentrations, etc. In practice, detailed identification of substructures is often challenging with ground-based data, where limited spatial resolution and source density hinder precise separation of overlapping halo components.

In this study, we revisit the Bullet Cluster to address the long-standing uncertainty in its mass ratio and merger configuration that has persisted for more than two decades. Our analysis combines James Webb Space Telescope (JWST) NIRCam and Dark Energy Camera (DECam) imaging to overcome the limitations of previous lensing studies. The JWST data provide an unprecedented background source density of $\mytilde400$~arcmin$^{-2}$,  $\mytilde4.5$ times higher than that of HST, enabling the identification and determination of 
three distinct halos.
The wide-field DECam coverage constrains the virial scale mass without extrapolation, while incorporating model-independent SL constraints into the WL framework anchors the mass profile and mitigates model bias with the SL mass serving as an effective proxy for concentration.
Through this integrated approach,
we demonstrate that the Bullet Cluster is a $\mytilde$10:1 minor merger, consistent with the mass ratios required by hydrodynamical simulations to reproduce its observed properties, thereby providing a unified picture of this archetypal merging system.

The paper is organized as follows. In \textsection{\ref{sec:obs}}, we describe our JWST and DECam observations and data reduction. \textsection{\ref{sec:ana}} details our WL analysis methodology, including point spread function (PSF) modeling, shape measurement, source selection, and source redshift estimation. \textsection{\ref{sec:res}} presents our mass reconstruction and estimation results, including our joint WL+SL analysis approach using SL-derived projected masses as boundary conditions for three-halo NFW profile fitting. \textsection{\ref{sec:dis}} presents an extensive comparison with prior work and evaluates the robustness of our results through a series of systematic tests.
We conclude in \textsection{\ref{sec:con}}. Throughout, we adopt a flat $\Lambda$CDM cosmology with $H_0 = 70$~km~s$^{-1}$~Mpc$^{-1}$, $\Omega_M = 0.3$, and $\Omega_{\Lambda} = 0.7$. At the cluster redshift ($z = 0.296$), the plate scale is $4.413$~kpc~arcsec$^{-1}$, and all observational images are oriented with north up and east to the left.

\section{Observations}
\label{sec:obs}

\begin{figure*}
\centering
\includegraphics[width=0.9\textwidth]{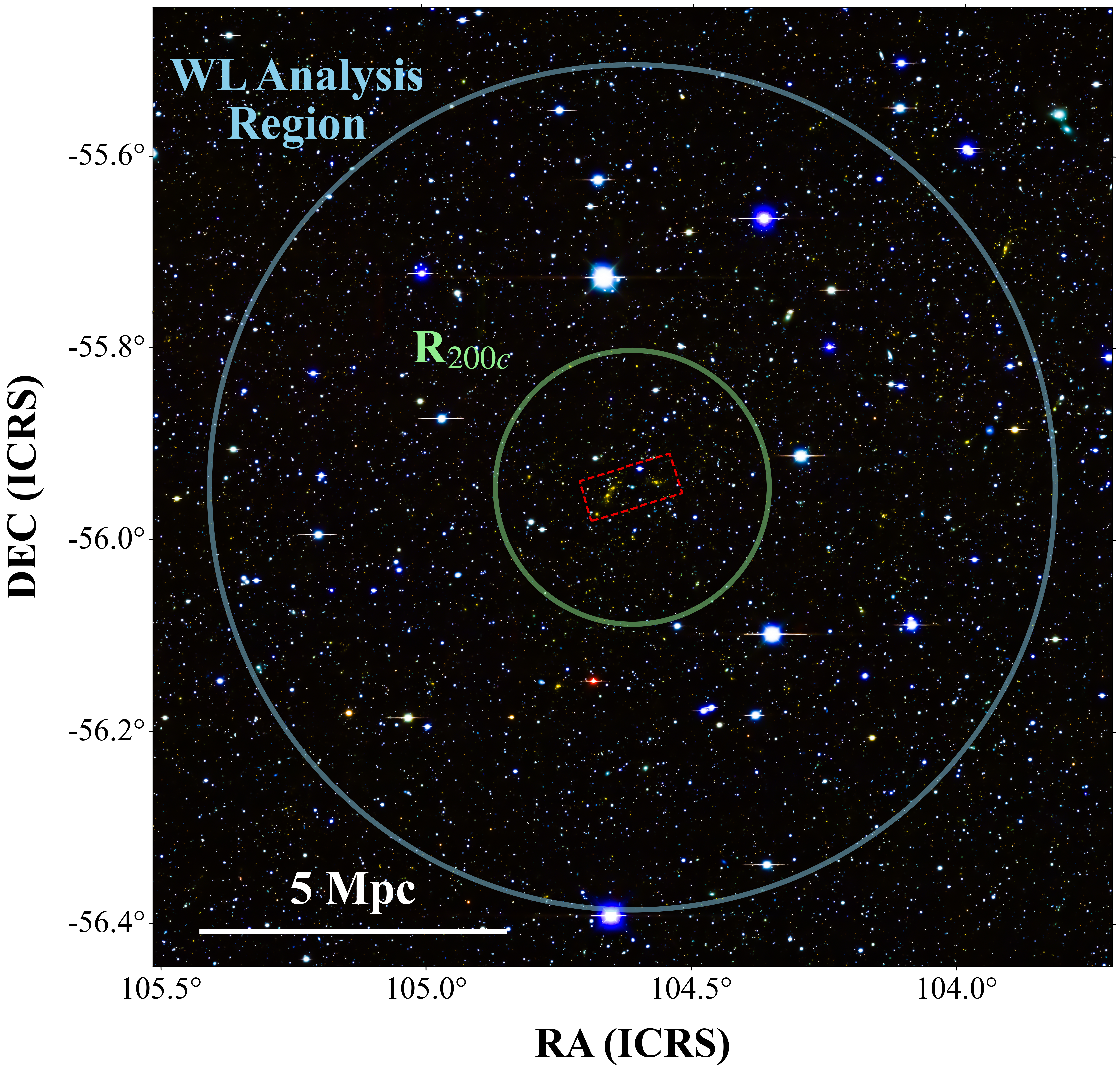} 
\caption{Wide-field view of the Bullet Cluster region. The background shows
our DECam $g$+$r$+$i$ color composite, covering
the $1^{\circ} \times 1^{\circ}$ ($\mytilde 16\text{ Mpc} \times 16\text{ Mpc}$ at $z=0.296$) region centered on the target. The red dashed rectangle marks the JWST/NIRCam F200W coverage ($\mytilde6\arcmin \times 2.5\arcmin$).
 The green circle
marks the
total system's virial radius ($R_{200c}^{\text{Total}}$ = 2.27~Mpc; \textsection{\ref{sec:bestfit}}), while the blue circle indicates the boundary of the DECam WL analysis region (7~Mpc radius) used for mass estimation.
}
\label{fig:1_coverage}
\end{figure*}

\subsection{JWST/NIRCam}
The JWST NIRCam imaging data from program GO-4598 (PI: M. Bradač) were used to perform 
a high-fidelity lensing analysis of the
Bullet Cluster's central region. The observations were obtained on 2025 January 20, covering $\mytilde6\arcmin \times 2.5\arcmin$ ($\mytilde1.6 \text{ Mpc} \times 0.7 \text{ Mpc}$ at $z = 0.296$).
All eight filters (F090W, F115W, F150W, F200W, F277W, F356W, F410M, and F444W) were used to identify strongly lensed multiple images, whereas WL shape measurements were performed only in F200W.

We
employed the standard JWST calibration pipeline \citep{bushouse_2024} with the parameter reference file map {\tt jwst\_1241.pmap}. We removed snowball artifacts arising from cosmic rays and pink noise using the algorithm described by \citet{Bagley_2023}. Additionally, we applied the third version of the wisp artifact templates to remove diffuse stray light features. For the final mosaic image, the
output pixel scale is set to $0\farcs02 ~\rm pixel^{-1}$ using a square kernel with {\tt pixfrac} = 0.8.
In \citet{Cha_2025}, we presented a two-dimensional projected mass distribution derived from both WL and SL data within the JWST field. In the present work, we focus on the WL constraints on the deprojected three-dimensional virial masses of the substructures, combining the JWST lensing data with the wide-field DECam observations described in the following subsection.

\subsection{Blanco 4m/DECam}
We use wide-field optical imaging data from the DECam on the Blanco 4 m telescope at Cerro Tololo Inter-American Observatory.
We combine Science Verification (SV) observations with subsequent programs: 2012B-0001, 2012B-0003, 2016A-0366, 2016A-0618, 2018A-0386, 2019A-0305, and 2019B-0323.
The retrieved DECam imaging data provide $\mytilde3.1$~deg$^{2}$ of
coverage,
constraining the Bullet Cluster's mass
beyond the JWST footprint (part of the entire DECam coverage shown in Figure~\ref{fig:1_coverage}).

The total exposure times are 6300, 1440, 2510, 2439, 1080, and 500 s for the $u$, $g$, $r$, $i$, $z$, and $Y$ filters, respectively. Although mosaics were constructed for all filters, only $g$, $r$, and $i$ were analyzed due to insufficient quality in the $u$, $z$, and $Y$ bands. These three filters were combined to create a detection image for {\tt SExtractor}, while the $g
 - i$ color, which brackets the 4000 $\mathring{\text{A}}$ break at $z = 0.296$,
 is used for source selection.
 Shape measurements were performed exclusively on the $i$-band images, which exhibit the best PSF model quality (\textsection{\ref{sec:psf}}).

Our initial attempt to process the DECam data using the LSST pipeline did not yield WL-quality images. We therefore performed a custom reduction to eliminate artifacts (e.g., scratch-like streaks, triangular corner patterns) and ensure WL-grade image fidelity.
Our reduction includes overscan subtraction and trimming, flat-fielding, and correction for amplifier gain differences. For flat-fielding, we created master sky flats using $\mytilde1430$ exposures from random sky positions observed within one month of the Bullet Cluster observations.

We then performed astrometric calibration and geometric distortion correction using SCAMP \citep{Bertin_2006} and SWarp \citep{Bertin_2010} with the Two Micron All Sky Survey as the reference catalog. We set {\tt POSITION\_MAXERR} = $7\arcmin$ to accommodate large pointing offsets in some SV files. Cosmic rays, saturated pixels, bad pixels, bleeding trails, and CCD edge artifacts (including tape marks) were masked before stacking exposures using a clipped mean algorithm \citep{Gruen_2014}. The resulting pixel scale is $\mytilde0\farcs26 ~\rm pixel^{-1}$.

\subsection{Source Detection and Photometry}
We performed source detection using {\tt SExtractor} \citep{Bertin_1996} in dual mode. For JWST, we created a detection image by inverse-variance-weighted averaging of eight NIRCam filters (F090W, F115W, F150W, F200W, F277W, F356W, F410M, and F444W). For DECam, we combined $g$, $r$, and $i$ filters. We set the detection parameters to require at least 5 connected pixels above $2\sigma$ ($1.5\sigma$) for DECam (JWST).

However, JWST member galaxies, particularly brightest cluster galaxies (BCGs), require special treatment due to their large angular extent relative to JWST's small pixel scale. Therefore, we applied an additional {\tt SExtractor} run with parameters optimized for member galaxies, requiring at least 50 connected pixels above $100\sigma$ and adjusting the background parameters to prevent fragmentation. This supplementary detection ensures proper identification and photometry of cluster members while maintaining our primary catalog of smaller background sources.

We measured colors using {\tt MAG\_ISO} and total magnitudes using {\tt MAG\_AUTO}. For JWST photometric calibration, we derived AB magnitude zeropoints
from the {\tt PHOTMJSR} and {\tt PIXAR\_SR} header keywords following standard JWST pipeline conventions. For DECam, we followed the methodology of \citet{Finner_2017}, matching instrumental magnitudes to the USNO-B1.0 catalog \citep{Monet_2003}. Since USNO-B1.0 provides only $B$ and $R$ magnitudes, we first transformed these to SDSS $g$ and $r$
magnitudes using the color transformations of \citet{Jester_2005} and then converted the results to the DECam filter system following
\citet{Abbott_2018}.

\subsection{Spectroscopic Data}
\label{sec:specdata}

We compiled spectroscopic redshifts from \citet{Richard_2021}, \citet{Puccetti_2020}, and \citet{Foex_2017} to identify cluster members within $0.296 \pm 0.014$ ($76,035^{+3006}_{-3055}$~km~s$^{-1}$). After removing duplicates (prioritizing sources in the order listed), we obtained 291 unique members: 23, 52, and 216 galaxies from these catalogs, respectively. We matched this spectroscopic catalog to our DECam and JWST catalogs using a $1\farcs0$ matching radius, yielding 277 members matched to DECam sources and 75 matched to JWST sources.

\section{Analysis}

\label{sec:ana}

\subsection{Basic WL Theory}

\label{sec:wl_theory}

We briefly review the basic WL formalism following established treatments in the literature \citep[e.g., ][]{Wright_2000, Bartelmann_2001, Schneider_2005, Hoekstra_2008, Hoekstra_2013, Kilbinger_2015, Mandelbaum_2018}. 
The lensing Jacobian matrix describes the transformation from the lens plane to the source plane coordinates:
\begin{equation}
    \mathcal{A}(\theta) = \frac{\partial\beta}{\partial\theta} = \begin{pmatrix}
    1-\kappa-\gamma_1 & -\gamma_2 \\
    -\gamma_2 & 1-\kappa+\gamma_1
    \end{pmatrix},
\end{equation} 
where $\beta$ and $\theta$ are the angular positions in the source and lens planes, respectively. The convergence $\kappa$ is defined as:
\begin{equation}
    \kappa(\theta) = \frac{\Sigma(D_l\theta)}{\Sigma_{cr}} \quad \text{with} \quad \Sigma_{cr} = \frac{c^2}{4\pi G} \frac{D_s}{D_l D_{ls}},
\end{equation}
where $\Sigma$ ($\Sigma_{cr}$) is the (critical) surface mass density. The shear components are $\gamma_1 = (\psi_{,xx} - \psi_{,yy})/2$ and $\gamma_2 = \psi_{,xy}$, where $\psi_{,x(y)}$ denotes the derivative of the lensing deflection potential $\psi$ with respect to $x(y)$.

The angular diameter distances $D_l$, $D_s$, and $D_{ls}$ represent observer-to-lens, observer-to-source, and lens-to-source distances, respectively. For sources at different redshifts, the lensing efficiency $\beta = D_{ls}/D_s$ quantifies the strength of the lensing signal, with increasing values for more distant sources.

The convergence $\kappa$ causes isotropic magnification while the shear $\gamma$ produces anisotropic distortion. In the WL regime ($\kappa \ll 1$), the reduced shear $g = \gamma / (1-\kappa)$ can be approximated as $g \approx \gamma$. Since we measure galaxy ellipticities $\epsilon$ rather than reduced shear directly, we rely on the statistical relationship $\langle \epsilon \rangle \approx \langle g \rangle$. This approximation holds because the expectation value of intrinsic ellipticities vanishes, $\langle e \rangle \approx 0$, when averaged over a sufficiently large galaxy sample. Thus, with $\langle \epsilon \rangle \approx \langle g \rangle \approx \langle \gamma \rangle$, we can reconstruct the convergence field from the observed ellipticity using the Kaiser-Squires inversion \citep{Kaiser_1993}:
\begin{equation}
    \kappa(\mathbf{x})=\frac{1}{\pi}\int_{\mathbb{R}^2}\mathcal{D}^*(\mathbf{x}-\mathbf{x'})\gamma(\mathbf{x'})d^2\mathbf{x'},
\end{equation}
where the convolution kernel $\mathcal{D}$ at position $(x_1, x_2)$ is:
\begin{equation}
    \mathcal{D}=-\frac{1}{(x_1-ix_2)^2}.
\end{equation}
Unless stated otherwise, we use {\tt FIATMAP} \citep{Fischer_1997, Wittman_2006} to generate
the WL mass map in this work.
{\tt FIATMAP} implements the Kaiser-Squires inversion in real space.

\subsection{PSF Modeling}
\label{sec:psf}

\begin{figure*}
\centering
\includegraphics[width=0.9\textwidth]{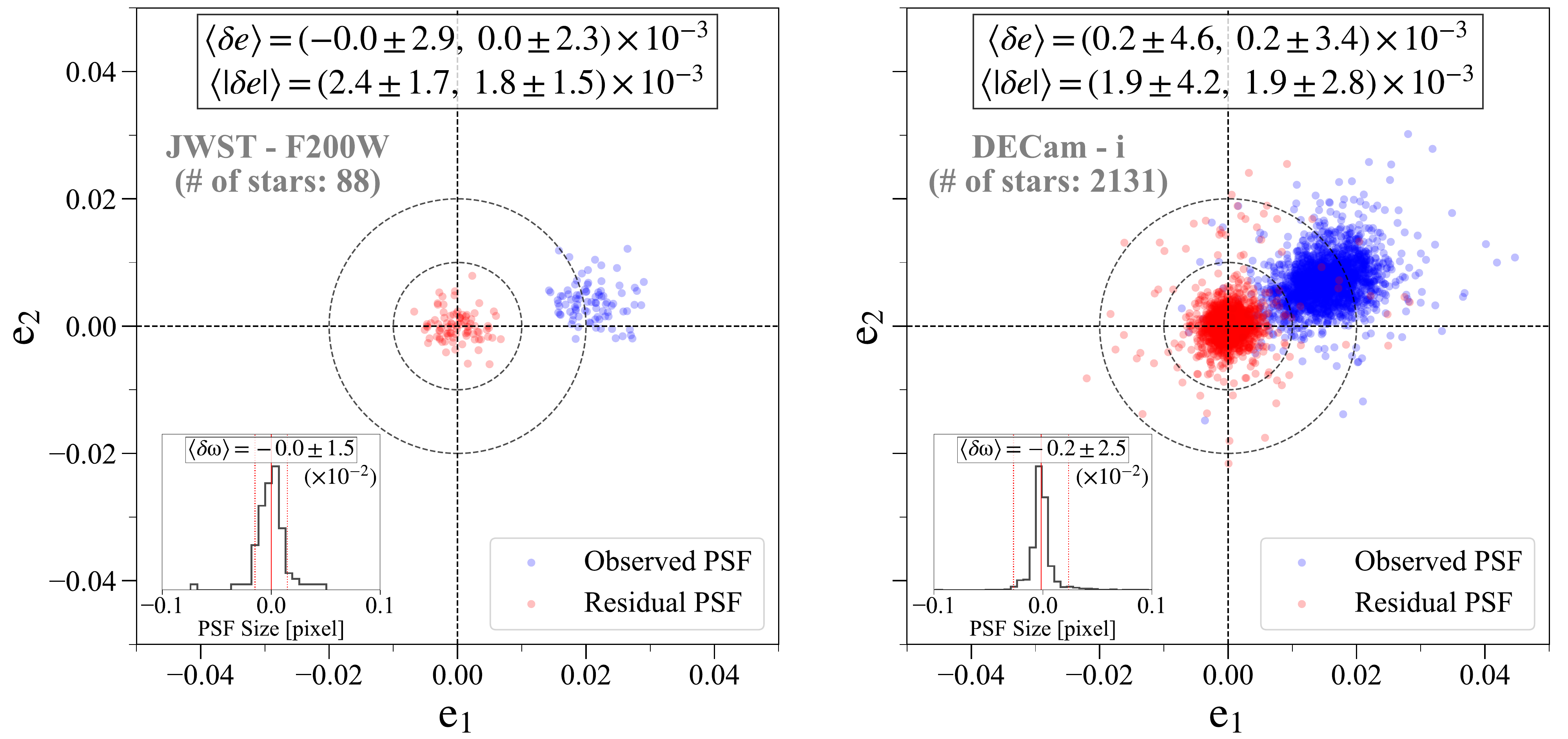} 
\caption{PSF correction quality for JWST F200W (left) and DECam $i$-band (right). Blue points show observed stellar ellipticities while red points show the ellipticity residuals after PSF correction (i.e., observed minus modeled ellipticity components). Residuals centered at (0,0) with small scatter demonstrate accurate and precise PSF modeling.
The lower-left inset in each panel displays PSF size residuals.}
\label{fig:3_psf}
\end{figure*}

\begin{figure*}
\centering
\includegraphics[width=0.9\textwidth]{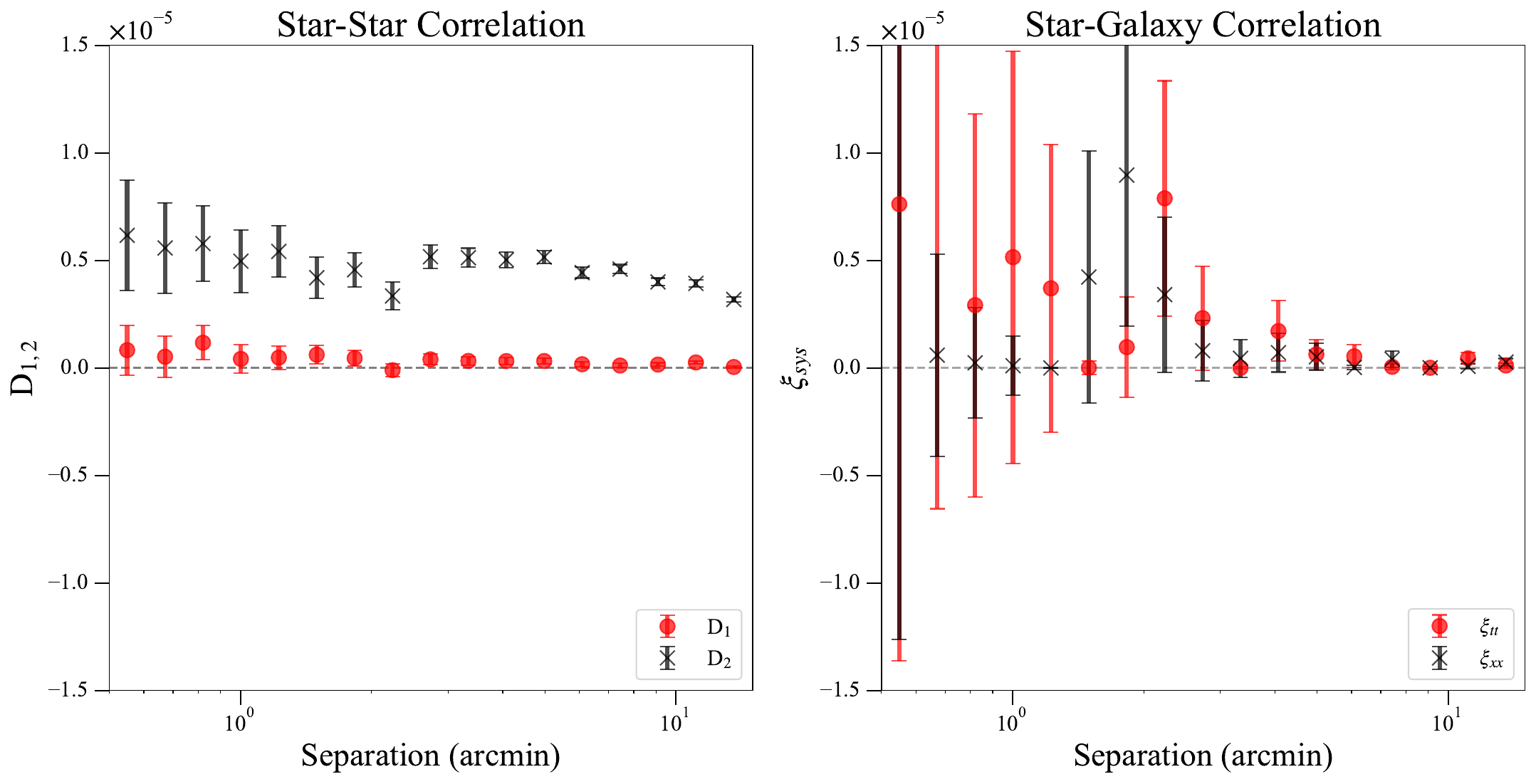} 
\caption{PSF model diagnostics for DECam $i$-band. \textit{Left}: $D_1$ and $D_2$ statistics \citep{Rowe_2010} showing auto-correlation of PSF model residuals ($D_1$; red circles) and cross-correlation between observed stellar ellipticities and residuals ($D_2$; black crosses). The $D_1$ ($D_2$) statistic remains below $10^{-6}$ ($10^{-5})$ across all angular scales, demonstrating exceptional PSF modeling accuracy. \textit{Right}: Star-galaxy correlation functions following \citet{Jee_2013}, measuring the normalized correlation $\xi_{\rm sys} = \langle e^* \gamma \rangle^2 / \langle e^* e^* \rangle$ decomposed into tangential ($\xi_{tt}$; red circles) and cross ($\xi_{\times\times}$; black crosses) components.
All correlation amplitudes remain below $10^{-5}$; notably, $\xi_{tt}$ and $\xi_{\times\times}$ fall below $10^{-6}$ once the separation exceeds $3'$.
These exceptionally low PSF-diagnostic levels demonstrate that any PSF-induced systematics are negligible in comparison to the WL signal.
}
\label{fig:3_psfcorr}
\end{figure*}

WL
measures mass through
subtle distortions in the 
ellipticities of background galaxies, making accurate removal of systematic effects paramount. Among these systematics, the PSF
is the most significant source of bias. PSF anisotropy
introduces coherent shape alignments that can mimic gravitational shear,
while even an isotropic PSF suppresses the lensing signal by enlarging galaxy images and reducing their observed ellipticities.
The quality of PSF correction directly sets the minimum resolvable galaxy size and the achievable source density, underscoring the central role of accurate PSF modeling in WL analyses.

We employ the PSF modeling technique using Principal Component Analysis (PCA) developed by \citet{Jee_2007}. This method has been successfully applied to various instruments including KPNO/CTIO Mosaic \citep{Jee_2013}, Subaru/Suprime-Cam \citep{Finner_2017}, HST/WFC3 \citep{Kim_2019}, DECam \citep{HyeongHan_2020}, Subaru/HSC \citep{Finner_2023A}, JWST/NIRCam \citep{Finner_2023B}, and Magellan/Megacam \citep{Ahn_2024}. We select ``good" stars (high signal-to-noise, isolated, unsaturated) across the field, apply subpixel shifts to center each star precisely, then extract principal components from mean-subtracted star stamps. We use 21 components empirically, which typically represent over 90\% of the data variance while reducing noise. For detailed methodology, see \citet{Jee_2007}.

For JWST, the stability of the PSF with respect to both time and position allows direct modeling from the mosaic image \citep{Finner_2023}.
We identified 88 ``good"
stars across the JWST field and applied the PCA method to create the PSF model.
In contrast, DECam requires individual CCD modeling due to spatially varying atmospheric and optical effects.
We performed PCA analysis on each CCD separately and then combined the results using the same weight-averaging scheme adopted for building the mosaic image.
Figure~\ref{fig:3_psf} illustrates that the resulting PSF models successfully recover both the ellipticity and size of the observed stellar profiles for both JWST and DECam.
For DECam, the $D_{1,2}$ diagnostics \citep{Rowe_2010} remain well controlled, with amplitudes far below $10^{-5}$ across all angular scales considered (left panel of Figure~\ref{fig:3_psfcorr}). The star--galaxy correlation (right panel of Figure~\ref{fig:3_psfcorr}) is similarly suppressed, reaching $\lesssim10^{-6}$, comfortably within the requirements for cluster WL analyses.

\subsection{Shape Measurement}
\label{sec:shape}

For JWST, we measure galaxy ellipticities by fitting PSF-convolved elliptical Gaussian functions to individual galaxy cutouts in the F200W mosaic. The fits are performed using {\tt MPFIT} \citep{Markwardt_2009}. The model consists of seven parameters: center position ($x, y$), variances ($\sigma_x^2, \sigma_y^2$), position angle, background, and peak value. We fix the center position and background from {\tt SExtractor} measurements and fit ($\sigma_x^2, \sigma_y^2$), position angle, and peak value. For DECam, we measure ellipticities from the $i$-band data, which shows the lowest
residual systematics in PSF diagnostics among the three filters. The $\chi^2$ minimization for shape measurement is given by:
\begin{equation}
    \chi^2 = \sum_{\text{pixels}}\left(\frac{I_i - G_i \otimes P_i}{\sigma_{\text{rms},i}}\right)^2,
\end{equation}
where $I$, $G$, $P$, and $\sigma_{\text{rms}}$ represent the observed cutout, elliptical Gaussian model, PSF, and noise, respectively, with subscript $i$ denoting pixel index.

The elliptical Gaussian fitting procedure systematically underestimates galaxy ellipticities due to two well-known effects: model bias, arising from the mismatch between simple Gaussian profiles and the true galaxy light distributions \citep{Jee_2014}, and noise bias, which results from the nonlinear propagation of pixel noise into ellipticity estimates.
To account for these effects, we apply multiplicative calibration factors of 1.11 and 1.07 to the JWST $e_1$ and $e_2$ components, respectively \citep{Finner_2023B}. For DECam, we adopt a multiplicative correction of 1.22 \citep{HyeongHan_2020}. All calibration factors were determined using the SFIT technique, which demonstrated superior performance in the GREAT3 challenge \citep{Mandelbaum_2015}.

\subsection{Source Selection and Redshift Estimation}
\label{sec:selection_redshift}

\begin{figure}
\centering
\includegraphics[width=\linewidth]{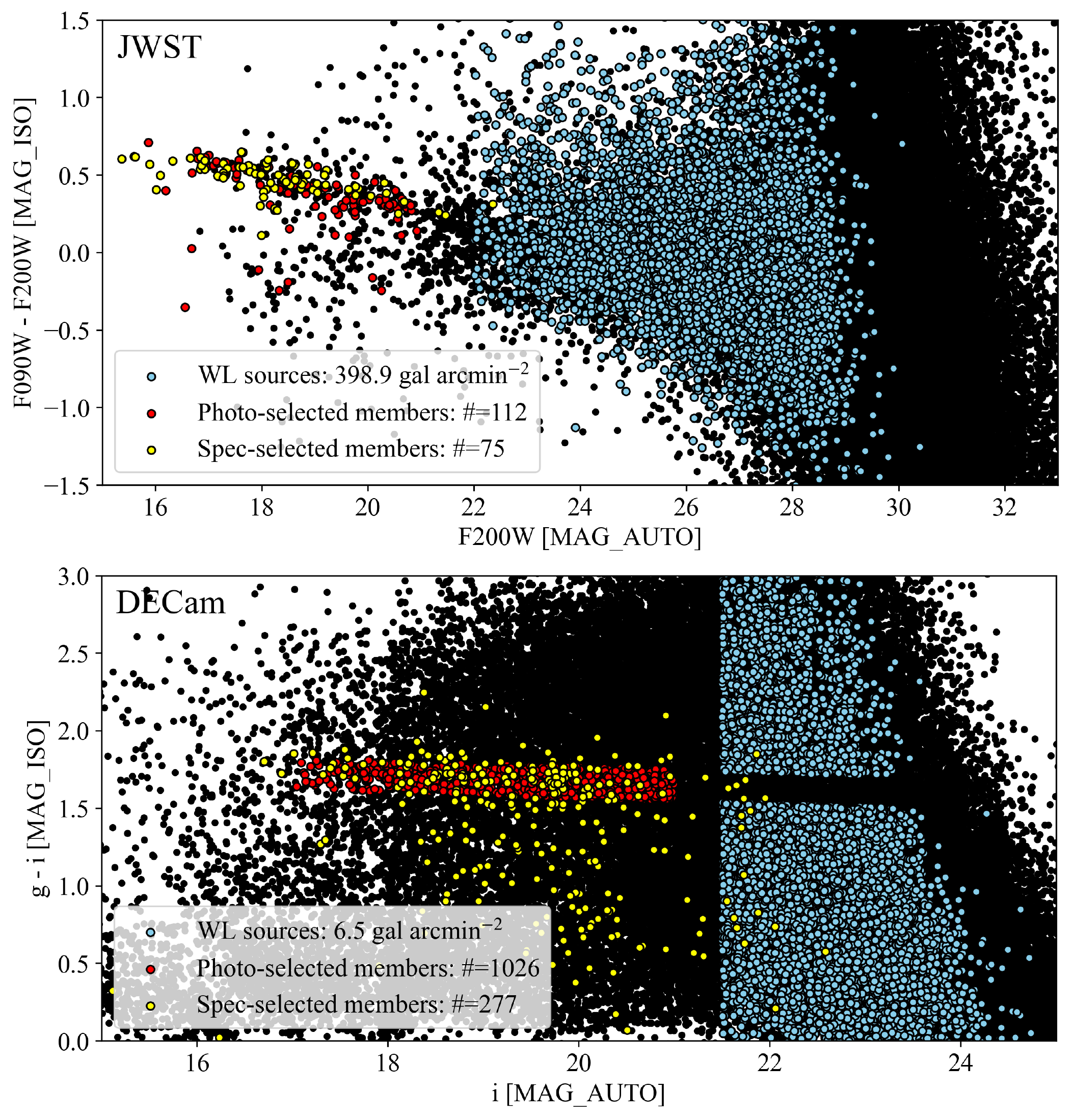} 
\caption{Source selection for JWST (upper) and DECam (lower) galaxies. Blue, red, and yellow points represent final WL sources, photometric member candidates, and spectroscopic members, respectively. For JWST, background source selection is based on photometric redshifts and a magnitude cut
(F200W $>22$).
For DECam, we determine the red sequence from a linear fit to the color-magnitude 
relation obtained from
spectroscopic members (\textsection{\ref{sec:specdata}}), adopting a color ($g-r$) width of $\pm$0.1 mag.
We selected sources fainter than $i>21.5$,  $\mytilde6$ mag below the brightest cluster members. 
To minimize potential contamination from faint red-sequence galaxies, we excluded objects lying within
$\pm$0.1 mag of the red-sequence color.
Foreground contamination in this selection was estimated using control fields and incorporated into the quantitative analysis of the shear signal (\textsection\ref{sec_decam_sources}).
}
\label{fig:3_selection}
\end{figure}

\begin{figure}
\centering
\includegraphics[width=\linewidth]{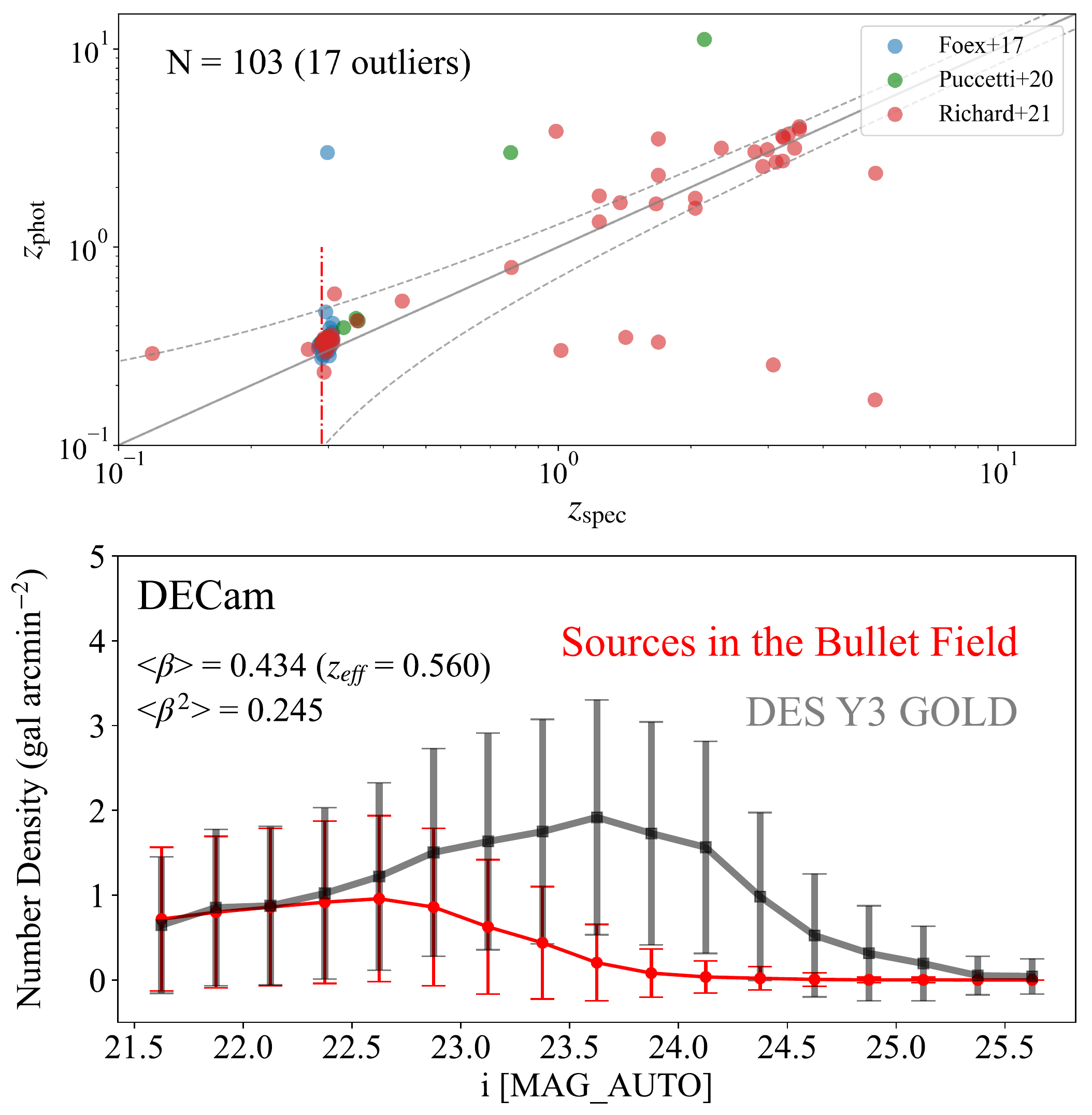} 
\caption{Spectroscopic and photometric redshift comparison for JWST sources and galaxy number density validation for DECam. \textit{Top}: Spectroscopic versus photometric redshifts for 103 JWST sources with spectroscopic data compiled from \citet{Foex_2017}, \citet{Puccetti_2020}, and \citet{Richard_2021} (\textsection{\ref{sec:specdata}}). The solid line indicates perfect agreement, while dashed lines mark 10\% deviation. Of the matched sources, 86 (83.5\%) show photometric redshifts within 10\% of spectroscopic values. The red vertical dashed line indicates the Bullet Cluster redshift ($z = 0.296$). \textit{Bottom}: Galaxy number density comparison between the sources in the Bullet Cluster field and DESY3GOLD control field as a function of magnitude for deriving lensing efficiency $\beta$. The excellent agreement between the two fields at the bright end ($\lesssim22.3$) indicates negligible member contamination in our source selection, while offsets at fainter magnitudes are due to differences in depth.}
\label{fig:3_redshift}
\end{figure}

We employ different
strategies for the JWST and DECam datasets due to the difference in the availability of individual photometric redshifts. 
For sources in the JWST field, we derive photometric redshifts using 13-band photometry from JWST/NIRCam (8 filters) and HST/ACS (5 filters).
In contrast, the DECam data offer only three optical bands ($g$, $r$, and $i$), which limits the reliability of individual photometric redshift estimates. We therefore adopt a red-sequence--based source selection followed by statistical redshift assignment using external control fields, for which high-fidelity photometric redshift catalogs are available.

\subsubsection{JWST Sources}

For JWST sources, we derive individual photometric redshifts from thirteen bands: eight JWST/NIRCam filters (F090W, F115W, F150W, F200W, F277W, F356W, F410M, and F444W) combined with five HST/ACS filters (F435W, F606W, F775W, F814W, and F850LP). We estimate photometric redshifts using {\tt EAzY} \citep{Brammer_2008} with the {\tt SFHz\_CORR} templates. This extensive multi-band coverage enables direct calculation of lensing efficiencies for individual sources. Validating against the spectroscopic sample (\textsection{\ref{sec:specdata}}), we find 83.5\% (86 out of 103) of matched sources have photometric redshifts within 10\% of spectroscopic values, defined as $|z_{\text{spec}} - z_{\text{phot}}|/(1+z_{\text{spec}}) < 0.10$ (Figure~\ref{fig:3_redshift}, top panel).

We select background galaxies based on three criteria: (1) photometric redshift uncertainties smaller than 25\%, (2) $1\sigma$ lower bounds (16th percentile) of the redshift probability distribution exceeding 0.35 (i.e., $z_{16} > 0.35$ corresponding to $z_{\text{min}} = 0.363$), which conservatively ensures line-of-sight velocity differences at least six times the velocity dispersion of typical massive merging clusters \citep{Golovich_2019, Finner_2025}, and (3) F200W magnitudes fainter than 22 to minimize cluster member and foreground contamination.

We apply shape quality criteria based on \citet{Jee_2013}: shape measurement error $<$ 0.3, ellipticity $<$ 0.9, semi-minor axis $>$ 0.4 pixels, and stable fitting convergence ({\tt STATUS} = 1 in {\tt MPFIT}; see \textsection{\ref{sec:shape}}). After applying size cut ({\tt FLUX\_RADIUS} $>$ 2.7 pixels), we manually discard problematic sources such as satellite trails, blended sources, and diffraction spikes, which is $\sim$4\% of the catalog (239 out of 6020 sources). In addition, all sources within 150 kpc of each halo center (the region used for SL anchoring) are excluded from the WL analysis, which also mitigates ICL contamination in the inner cluster: outside this radius, the ICL profile is sufficiently shallow to be absorbed into the local sky background estimated by SExtractor. The final JWST WL catalog reaches a source density of 398.9~arcmin$^{-2}$ (Figure~\ref{fig:3_selection}). This exceptionally high density reflects the telescope’s superior spatial resolution and depth.

\subsubsection{DECam Sources}
\label{sec_decam_sources}
For DECam, we employ the red-sequence method utilizing the 4000 $\mathring{\text{A}}$ break (redshifted to 5184 $\mathring{\text{A}}$ at $z = 0.296$), which falls between the $g$ and $i$ bands. We
determine the red-sequence
locus
from a linear fit to the color-magnitude distribution of spectroscopically confirmed members (\textsection{\ref{sec:specdata}}), adopting a color width of $\pm0.1$ magnitudes from the best-fit line for galaxies brighter than 21 mag. Background sources are identified as galaxies bluer or redder than the red sequence with magnitudes fainter than 21.5 mag to minimize member contamination while retaining sufficient source density. We impose the same shape-quality requirement used for JWST source selection, size cut ({\tt FLUX\_RADIUS} $>$ 2.3 pixels), and the manual discarding process. The final DECam source density is 6.5~arcmin$^{-2}$ (Figure~\ref{fig:3_selection}).

For redshift assignment in the absence of individual photometric redshift, we adopt a statistical approach based on the Dark Energy Survey Year 3 GOLD catalog (DESY3GOLD; \citealt{Sevilla-Noarbe_2021}), restricted to the GOODS-S footprint, which is widely used as a control field in WL analyses. Because this field contains no massive clusters, its galaxy population consists primarily of field galaxies and therefore provides an appropriate reference redshift distribution for our background sources.

We computed the DECam lensing efficiencies using the redshift distribution from the control field. To account for differences in observational depth, we weight the redshift distribution by the source's number count ratio between the Bullet Cluster and control field in each magnitude bin ($\Delta m = 0.25$ mag; Figure~\ref{fig:3_redshift}, bottom panel). The representative lensing efficiency for each bin is:
\begin{equation}
    \beta = \langle \max(0, D_{ls}/D_s) \rangle.
\end{equation}
Galaxies with redshifts below the cluster redshift are assigned  $\beta$ = 0 because they are unlensed.

This process yields a lensing efficiency of $\langle\beta\rangle = 0.434$ with effective source plane redshift $z_\text{eff} = 0.560$. We calculate $\langle \beta^2 \rangle$ = 0.245 to account for the width of the source redshift distribution and apply the first-order correction \citep{Seitz_1997}:
\begin{equation}
    g' = \left[1 + \left(\frac{\langle\beta^2\rangle}{\langle\beta\rangle^2} - 1\right)\kappa\right]g,
\end{equation}
where $g'$ is the corrected reduced shear. This correction prevents mass overestimation when representing a distributed source population with a single efficiency value.

The Bullet Cluster field exhibits galaxy number densities remarkably consistent with the control field (Figure~\ref{fig:3_redshift}, bottom panel), demonstrating negligible member contamination in our source catalog. This excellent agreement validates our source selection criteria. The systematically higher number densities observed in the control field at fainter magnitudes ($i \gtrsim$ 22.3) reflect its greater depth.

\section{Results}
\label{sec:res}

\subsection{Mass Reconstruction}
\label{sec:recon}

\subsubsection{JWST High-Resolution Mass Map}

\begin{figure*}
\centering
\includegraphics[width=0.9\textwidth]{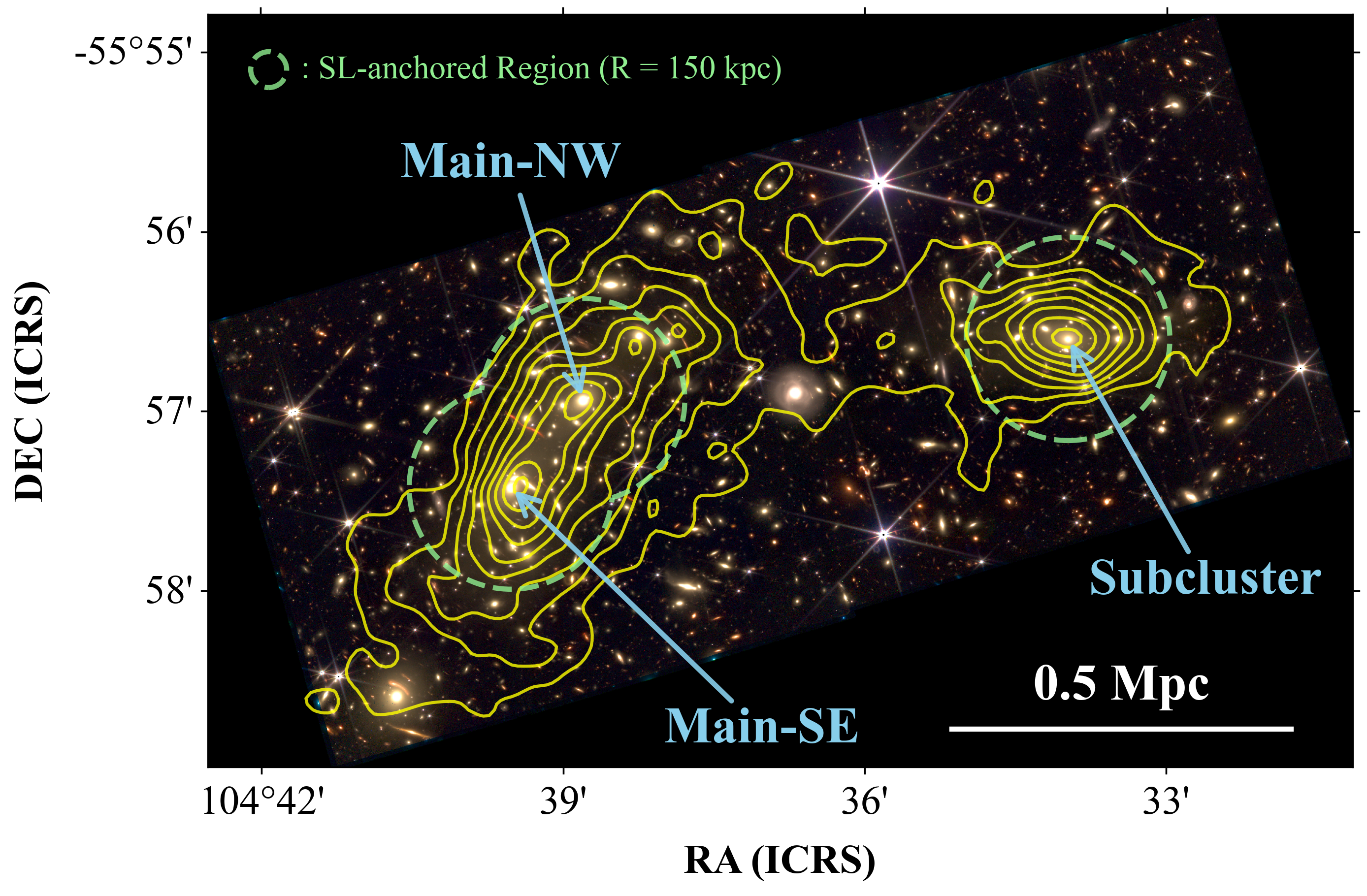} 
\caption{
WL+SL mass reconstruction of the Bullet Cluster from JWST/NIRCam data using the {\tt MARS} algorithm \citep{Cha_2022}.
We reproduced the mass contours in \citet{Cha_2025}, ranging from
$\kappa=0.15$ to 1.35 in steps of 0.15.
Green dashed circles indicate the  SL-anchored regions (150~kpc radii from the three BCGs) used as boundary conditions for WL fitting in the current study. Source galaxies within these regions are excluded from the WL analysis.}
\label{fig:4_recon_jwst}
\end{figure*}

Our WL+SL mass reconstruction from JWST/NIRCam, presented in \citet{Cha_2025}, provides the highest-resolution view of the Bullet Cluster's central mass distribution to date, combining 146 SL constraints with a WL source density of $\mytilde400$~arcmin$^{-2}$.
The resulting mass map yields a stable and well-resolved determination of the inner convergence field, including the locations and amplitudes of the two primary mass peaks in the main cluster. In the present work, we adopt this reconstruction to anchor the central region of our wide-field WL analysis, thereby ensuring a consistent mass normalization.

Figure~\ref{fig:4_recon_jwst} illustrates the anchored regions. These regions are defined based on the three mass concentrations identified in the Bullet Cluster system: two associated with the main cluster (labeled Main-NW and Main-SE) and one associated with the subcluster (labeled Subcluster). Because the \citet{Cha_2025} reconstruction robustly constrains the convergence $\kappa$ within the SL regime ($\lesssim150$~kpc from each peak), we define three circular anchoring regions, as shown. The two main-cluster regions overlap owing to their proximity.

The projected mass enclosed within the anchoring region in the main cluster is $(1.37 \pm 0.12)\times10^{14}M_{\odot}$, while the corresponding value for the subcluster is $(5.79 \pm 0.67)\times10^{13}M_{\odot}$. These mass estimates are used as fixed constraints to anchor our wide-field WL fitting (\textsection\ref{sec:estim}). We caution that a naive comparison of the projected masses within the SL-anchored regions could misleadingly suggest that the system is undergoing a $\mytilde$2:1 major merger. Treating these localized projected masses as direct proxies for the total halo masses would lead to an incorrect characterization of the merger scenario.

\subsubsection{DECam Wide-Field Mass Map}

\begin{figure*}
\centering
\includegraphics[width=0.9\textwidth]{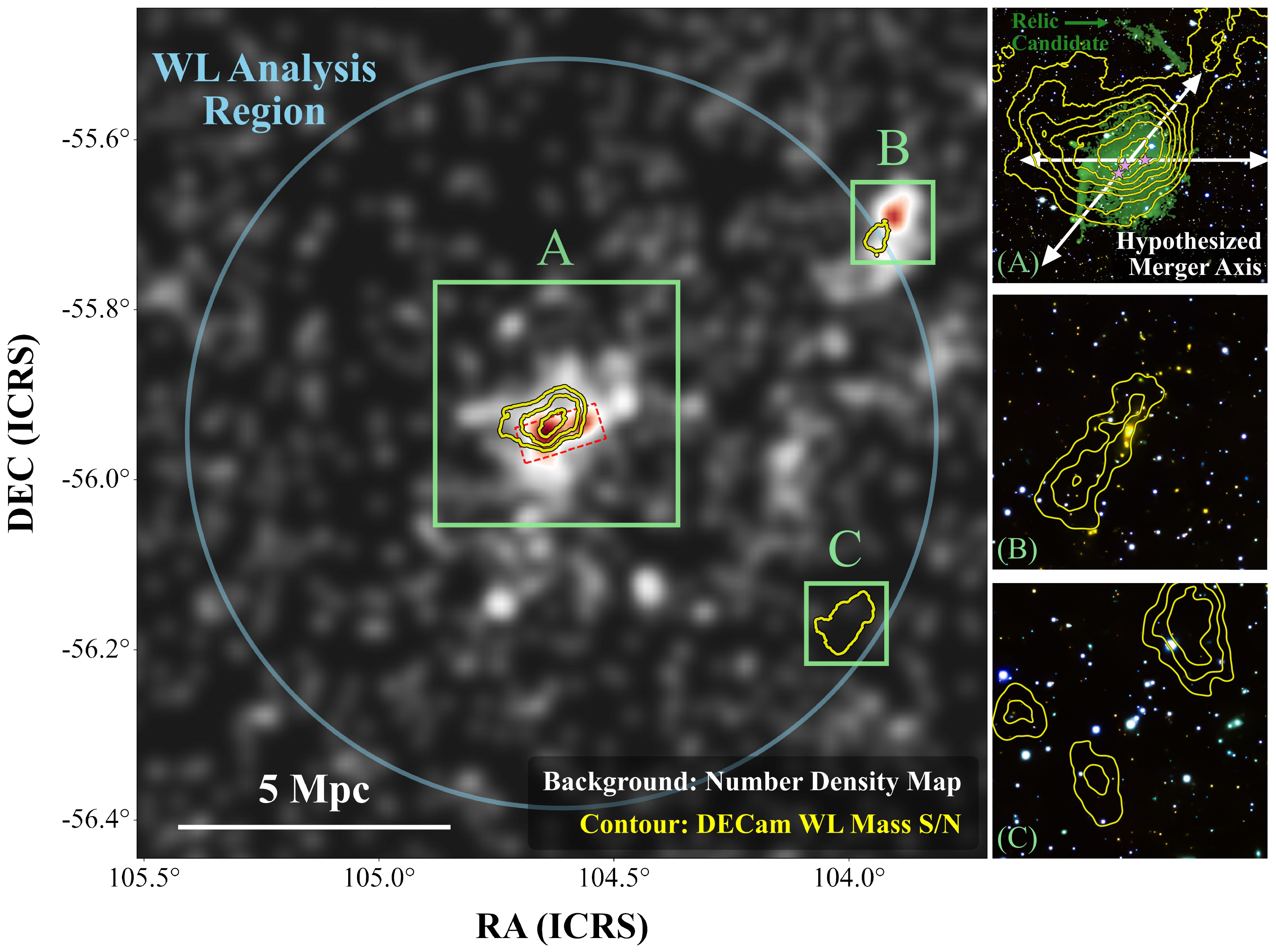} 
\caption{Wide-field mass reconstruction of the Bullet Cluster from DECam data, covering $1^{\circ} \times 1^{\circ}$. 
\textit{Left:} The background grayscale represents the galaxy number density map derived from our red-sequence catalog (Figure~\ref{fig:3_selection}).
The red dashed rectangle marks the JWST/NIRCam F200W coverage ($\mytilde6\arcmin \times 2.5\arcmin$), while the blue circle indicates the boundary of the DECam WL analysis region (7~Mpc radius) used for mass estimation.
Yellow contours indicate the WL mass reconstruction, plotted as signal-to-noise (S/N) levels from 3 to 4 in increments of 0.5. The S/N values are computed by dividing the mass map by the rms map obtained from 1000 bootstrap realizations.
We identified three regions (labeled as A, B, and C) where the S/N levels exceed $3\sigma$ and displayed their zoom-in views in panels A-C, where the background is DECam $g$+$r$+$i$ color image.
($A$) The Bullet Cluster field enclosing the $\mytilde4.5 \text{ Mpc} \times 4.5 \text{ Mpc}$ region. The three pink stars indicate the positions of the BCGs associated with the three mass peaks revealed in the JWST mass map.
Intensities in green represent the radio emission from MeerKAT \citep{Knowles_2022}. The two arrows illustrate two hypothesized merger axes (see text).
($B$) Shear-selected cluster candidate detected $\mytilde7.5$~Mpc northwest of the Bullet Cluster. The field size is $1.5 \text{ Mpc} \times 1.5 \text{ Mpc}$. The contours, which coincide with the galaxy overdensity, range from S/N = 2 to 3 in increments of 0.5 and are generated using a smoothing kernel that is 20\% smaller than the one adopted in the main panel.
($C$)  The contours are drawn using the same scheme as in the Region B panel. No significant galaxy overdensity is detected in the vicinity of these contours.
}
\label{fig:4_recon_decam}
\end{figure*}

Figure~\ref{fig:4_recon_decam} presents the wide-field mass reconstruction over
the same area shown in Figure \ref{fig:1_coverage}. 
We identified three high-significance regions in the DECam WL map.
The strongest signal (Region A) corresponds to the Bullet Cluster, with a peak significance of $S/N\sim4$. The relatively low WL source density (6.5~arcmin$^{-2}$) does not resolve the substructures revealed in the JWST result. Comparison with the MeerKAT radio data \citep{Knowles_2022} shows that the eastern radio relic lies approximately perpendicular to (and thus consistent with) the hypothesized E-W merger axis (horizontal arrow in panel A). We propose a secondary merger axis (NW-SE arrow in panel A) connecting the two substructures identified with JWST in the main-cluster region (Figure~\ref{fig:4_recon_jwst}). This secondary merger axis, tilted $\mytilde50^{\circ}$ with respect to the primary (E-W) merger axis, is also roughly perpendicular to the radio relic candidate \citep{Sikhosana_2023} $\mytilde2$~Mpc northwest of the Bullet Cluster. Based on the hypothesis that this is indeed a radio relic from this secondary merger, \cite{Lee_2025} estimated that the pericenter passage happened $\mytilde1.1$ Gyr ago.

A second mass overdensity (Region B), located roughly $\mytilde7.5$~Mpc northwest of the Bullet Cluster, shows good spatial coincidence with an overdensity of galaxies whose colors are consistent with the Bullet Cluster red sequence. This mass clump has not been reported in the previous DECam WL study \citep{Melchior_2015}. The peak significance of this cluster candidate is $S/N\sim3$.

For the third WL overdensity (Region C), we could not identify any galaxy overdensity. This feature could represent a higher-redshift cluster whose galaxy population is too faint for our current DECam data, or it may simply be due to chance alignment. We find no evidence of elevated systematics in this region.

Aside from the Bullet Cluster itself and the two structures in Regions B and C near the field boundary, we detect no additional significant WL features across the DECam field. This supports modeling the system as a composition of three halos when estimating the total mass of the Bullet Cluster system.

\subsection{Mass Estimation}
\label{sec:estim}

\subsubsection{Three-Halo Modeling Framework and WL+SL Mass Estimation}
The JWST reconstruction (Figure~\ref{fig:4_recon_jwst}) reveals three distinct mass peaks: two associated with the main cluster and one with the subcluster. Furthermore, the wide-field DECam result shows no additional significant structures within the $r=7$~Mpc boundary, aside from two weak overdensities near the field edge. These findings strongly motivate modeling the Bullet Cluster’s WL signal using three distinct halos.

We center each halo at the position of its corresponding BCG, as JWST WL+SL mass peaks are well aligned with the BCGs. Each halo is modeled with an NFW profile without imposing an $M$--$c$ relation. Although an $M$--$c$ relation is often adopted in standard WL analyses to reduce the intrinsic degeneracy between mass and concentration, merging clusters are known to deviate substantially from this relation \citep{Finner_2025}. Their dynamical disturbance can alter halo structures enough to induce mass biases of up to $\mytilde60$\% when an $M$--$c$ prior is enforced \citep{Lee_2023}.

We employ Markov Chain Monte Carlo (MCMC) sampling using {\tt emcee} \citep{Foreman-Mackey_2013} with 100 walkers and 10,000 steps per walker, discarding the initial 10\% as burn-in. To verify chain convergence, we performed additional tests with 1,000 walkers or 100,000 steps, yielding consistent posterior distributions and confirming the robustness of our configuration. We adopt uniform priors on mass ($10^{13} < M_{200c}/M_{\odot} < 10^{16}$) and concentration ($1 < c_{200c} < 20$), where $M_{200c}$ denotes the mass within a radius where the mean enclosed density equals 200 times the critical density at the cluster redshift. These settings are used consistently throughout our analyses, including the systematic tests in \textsection\ref{sec:model_comparison}.

The WL log-likelihood is:
\begin{equation}
\chi^{2} = \sum_{i=1}^{N_{\text{gal}}} \sum_{j=1}^{2}\frac{(g_{i,j}^{\text{model}}-\epsilon_{i,j}^{\text{obs}})^2}{\sigma_{s}^2 + \sigma_{m,i}^2},
\end{equation}
where $g_{i,j}^{\text{model}}$ represents the reduced shear's $j$-th component from the superposition of three NFW profiles evaluated at galaxy $i$'s position, $\epsilon_{i,j}^{\text{obs}}$ is the observed ellipticity, $\sigma_s$ is the intrinsic shape noise, and $\sigma_{m,i}$ is the measurement uncertainty. We note that the WL data are not binned; the model is fit directly to the shapes of individual source galaxies.

As noted earlier, we perform a joint WL+SL analysis by incorporating the projected masses constrained by the JWST  reconstruction from \citet{Cha_2025} as boundary conditions in our WL fitting. 
This approach mitigates the model bias highlighted by \citet{Lee_2023}, who showed that halo contraction during core passage in merging clusters can substantially increase the inferred concentration, leading to WL mass biases of up to $\mytilde60$\%. \citet{Lee_2023} also suggested that such contraction-induced biases can be significantly reduced when reliable central-mass constraints, such as those provided by SL measurements, are incorporated into the modeling.

To implement this scheme, we require our model not only to fit the galaxy ellipticities but also to reproduce the projected masses in the anchoring regions (Figure~\ref{fig:4_recon_jwst}). At each MCMC step, we compute the total projected mass within these regions by summing the contributions from the three NFW halos. We discard any proposed MCMC sample whose projected masses are inconsistent with the anchoring values: $(1.37 \pm 0.12)\times10^{14}M_{\odot}$ for the main cluster and $(5.79 \pm 0.67)\times10^{13} M_{\odot}$ for the subcluster.

Although the masses within the anchoring regions in \citet{Cha_2025} are primarily constrained by the SL data, a small fraction of WL information from JWST also contributes. To avoid double counting these constraints, we exclude JWST WL source galaxies inside the anchoring regions from our composite WL catalog, which combines JWST and DECam data.

\subsubsection{Impact of Uncorrelated Large-Scale Structures}
\label{sec:uncorr_LSS}
In addition to the Bullet Cluster itself, the wide-field DECam WL signal inevitably contains contributions from uncorrelated large-scale structure (LSS) along the line-of-sight \citep{Vecchi_2025}. These LSS fluctuations add coherent shear patterns that can affect the inferred masses. Because our mass constraints rely on fitting the full shear field over a circular region of radius 7~Mpc, we need  to incorporate the impact of LSS into our total error budget rather than attributing all shear solely to the cluster halos.

We quantify projection effects from LSS using \textsc{kappaTNG} mock WL maps \citep{Osato_2021}, selecting realizations at $z_s = 0.552$ (closest to DECam's effective source redshift $z_s = 0.560$). We generate 1000 mock shear realizations at our DECam source galaxy positions and add their reduced shear components directly to the observed values. For each realization, we repeat our complete mass estimation procedure. The LSS-induced uncertainties are calculated as the standard deviation of median values across all realizations.

\subsubsection{Impact of Correlated Large-Scale Structures }
\label{sec:corr_LSS}

We described in \textsection\ref{sec:uncorr_LSS} how we estimated the additional mass uncertainties arising from uncorrelated LSS. To evaluate the potential influence of correlated LSS at the Bullet Cluster redshift, we tested the impact by adding the two-halo term following the prescription of \citet{Wu_2019}. This term accounts for the contribution of surrounding large-scale structure to the outer halo density profile and is commonly included in stacked cluster-lensing analyses; here, we quantify its impact on individual-cluster mass inferences.

Relative to our fiducial results (which do not include this correction), the two main-cluster halo masses shift lower by only $\mytilde$1--3\%, while the subcluster mass increases by $\mytilde$0.7\%. The inferred mass ratios remain virtually unchanged, and all shifts lie comfortably within the $1\sigma$ uncertainties of the uncorrected posteriors. These small deviations indicate that the correlated LSS contribution is subdominant at the radii where our WL+SL constraints dominate. Because correlated LSS corrections are typically derived and applied in stacked cluster analyses, and because their sign and magnitude are not reliably predictable for individual systems, we report our fiducial results without incorporating this correction, while noting its minimal impact in the present case.

\subsubsection{Best-fit Results from the Fiducial Model}
\label{sec:bestfit}

\begin{figure*}
\centering
\includegraphics[width=0.9\textwidth]{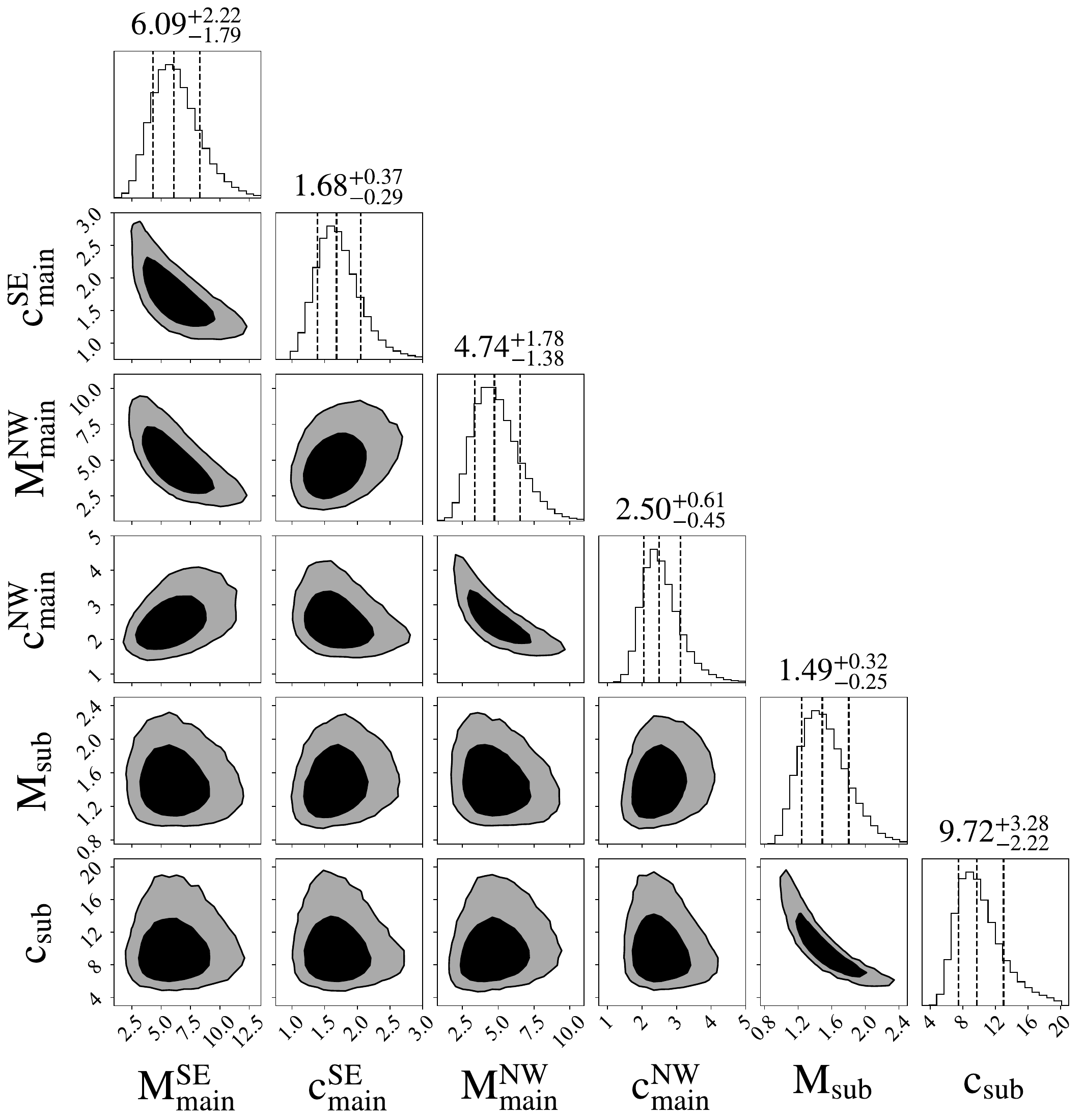} 
\caption{MCMC posterior distributions for our fiducial case ({\tt 3cS}; this notation from Table~\ref{table:notation}). All masses are in units of $10^{14}M_{\odot}$. The corner plot shows marginalized $M_{200c}$ and $c_{200c}$ parameters of all three components: Main-SE, Main-NW, and Subcluster, with center positions fixed at respective BCGs during sampling. Contours represent 68\% and 95\% confidence regions.}
\label{fig:5_estim_corner}
\end{figure*}

\let\oldtablecomments\tablecomments

% Left-aligned comments at full table* width
\renewcommand{\tablecomments}[1]{%
  \par\begingroup
  \footnotesize
  \parbox{\columnwidth}{\raggedright #1}%
  \endgroup
}

\begin{table}
\caption{Model Configuration Notation}
\centering
\small
\setlength{\tabcolsep}{8pt}
\renewcommand{\arraystretch}{1.2}
\begin{tabular}{cl}
\hline\hline
Symbol & Description \\
\hline
2 & Two-halo model \\
\textbf{3} & \textbf{Three-halo model} \\
\hline
M & With $M$--$c$ relation \\
\textbf{c} & \textbf{Without $M$--$c$ relation} \\
\hline
\textbf{S} & \textbf{with SL constraints (i.e., WL + SL)} \\
L & without SL constraints (i.e., WL only) \\
\hline
\end{tabular}
\tablecomments{Description of our configuration notation in the format (Number)($M$--$c$)(SL). Three-halo models (3) place NFW halos at each BCG position: Main-SE, Main-NW, and Sub. Two-halo models (2) place halos at the midpoint between main BCGs and at the subcluster BCG. SL constraints (S) require the WL NFW models to match the SL-projected masses within SL-anchored regions, accounting for overlapping contributions in the main-cluster region. The $M$--$c$ relation (M) follows \citet{Diemer_2019}. We note the configuration for our fiducial model shown in bold text.}
\label{table:notation}
\end{table}

\let\oldtablecomments\tablecomments

\renewcommand{\tablecomments}[1]{%
  \par\begingroup
  \footnotesize
  \parbox{\textwidth}{\raggedright #1}%
  \endgroup
}

\begin{table*}
\caption{Three-Halo NFW Profile Fitting Results}

\centering
\small 
\setlength{\tabcolsep}{4pt} 
\renewcommand{\arraystretch}{1.2}
\begin{tabular}{lcccccc}
\hline\hline
Config. & \multicolumn{2}{c}{Main-SE} & \multicolumn{2}{c}{Main-NW} & \multicolumn{2}{c}{Subcluster} \\
\cline{2-3}\cline{4-5}\cline{6-7}
& $M_{200c}$ & $c_{200c}$ & $M_{200c}$ & $c_{200c}$ & $M_{200c}$ & $c_{200c}$ \\
\hline
$\bm{3cS}$ &
$\bm{6.09^{+2.22}_{-1.79}\pm0.15}$ & $\bm{1.68^{+0.37}_{-0.29}\pm0.03}$ &
$\bm{4.74^{+1.78}_{-1.38}\pm0.04}$ & $\bm{2.50^{+0.61}_{-0.45}\pm0.03}$ &
$\bm{1.49^{+0.32}_{-0.25}\pm0.01}$ & $\bm{9.72^{+3.28}_{-2.22}\pm0.07}$ \\
$3cL$ &
$6.27^{+2.52}_{-1.96}\pm0.17$ & $1.63^{+0.39}_{-0.30}\pm0.04$ &
$4.84^{+2.03}_{-1.48}\pm0.07$ & $2.43^{+0.65}_{-0.47}\pm0.04$ &
$1.54^{+0.33}_{-0.26}\pm0.02$ & $9.63^{+3.22}_{-2.17}\pm0.07$ \\
\hline
\end{tabular}
\tablecomments{Results from MCMC analysis of three-halo NFW profile fitting without $M$--$c$ relation. All masses are in units of $10^{14}M_{\odot}$. Configuration notation follows Table~\ref{table:notation}. The asymmetric errors represent 1$\sigma$ marginalized uncertainties, while the $\pm$ term following each best-fit mass or concentration is the additional LSS-induced uncertainty derived in \textsection{\ref{sec:uncorr_LSS}}. Our fiducial model {\tt 3cS} is shown in bold.}
\label{table:4_main_results}
\end{table*}

Here, we focus on the results from our fiducial model, which employs three NFW halos without imposing an $M$--$c$ relation and incorporates the SL constraints. We refer to this model as {\tt 3cS}. Table~\ref{table:notation} summarizes the labeling scheme for non-fiducial models explored in this study.

Figure~\ref{fig:5_estim_corner} shows the MCMC posterior distributions for our fiducial {\tt 3cS} model. All six parameters describing the three NFW halos are well constrained. The corresponding best-fit parameter values, including uncertainties from LSS contributions, are summarized in Table~\ref{table:4_main_results}. We obtain masses of $M_{200c}^{\mathrm{Main\text{-}SE}} = (6.09^{+2.22}_{-1.79} \pm 0.15)\times10^{14}M_{\odot}$ and $M_{200c}^{\mathrm{Main\text{-}NW}} = (4.74^{+1.78}_{-1.38} \pm 0.04)\times10^{14}M_{\odot}$ for the two main-cluster components, and $M_{200c}^{\mathrm{Sub}} = (1.49^{+0.32}_{-0.25} \pm 0.01)\times10^{14}M_{\odot}$ for the subcluster.
The result shows that the two halos in the main cluster have comparable masses whereas the subcluster's mass is lower by a factor of 3--4.

To derive the total mass in the main cluster, we construct a three-dimensional density grid from the three NFW halos in our best-fit model, assuming that differences in their line-of-sight positions are negligible.
Starting from the midpoint between the two main BCGs, we perform spherical integration outward until the mean enclosed density equals 200$\rho_c$. This calculation is performed for every MCMC sample after burn-in. From the resulting posterior distributions, we obtain $M_{200c}^{\text{Main}} = 15.11^{+2.48}_{-2.10} \times 10^{14}M_{\odot}$. The corresponding mass ratio, derived from the MCMC posterior to fully account for parameter degeneracies, is $M_{200c}^{\text{Main}}/M_{200c}^{\text{Sub}} = 10.14^{+3.22}_{-2.47}$, which strongly favors a minor-merger scenario.

In a similar manner, we derive the total mass of the Bullet Cluster system by combining the contributions from all three halos and adopting the midpoint between the main-cluster and subcluster centers as the reference. We obtain $M_{200c}^{\text{Total}} = 17.99^{+2.56}_{-2.18} \times 10^{14}M_{\odot}$ with the corresponding virial radius $R_{200c}^{\text{Total}} = 2.27\pm0.10$~Mpc.

\section{Discussion}
\label{sec:dis}

\subsection{Systematic Discrepancies in Previous Bullet Cluster Mass Estimates}
\label{5_literaturecomparison}

\begin{figure*}
\centering
\includegraphics[width=\textwidth]{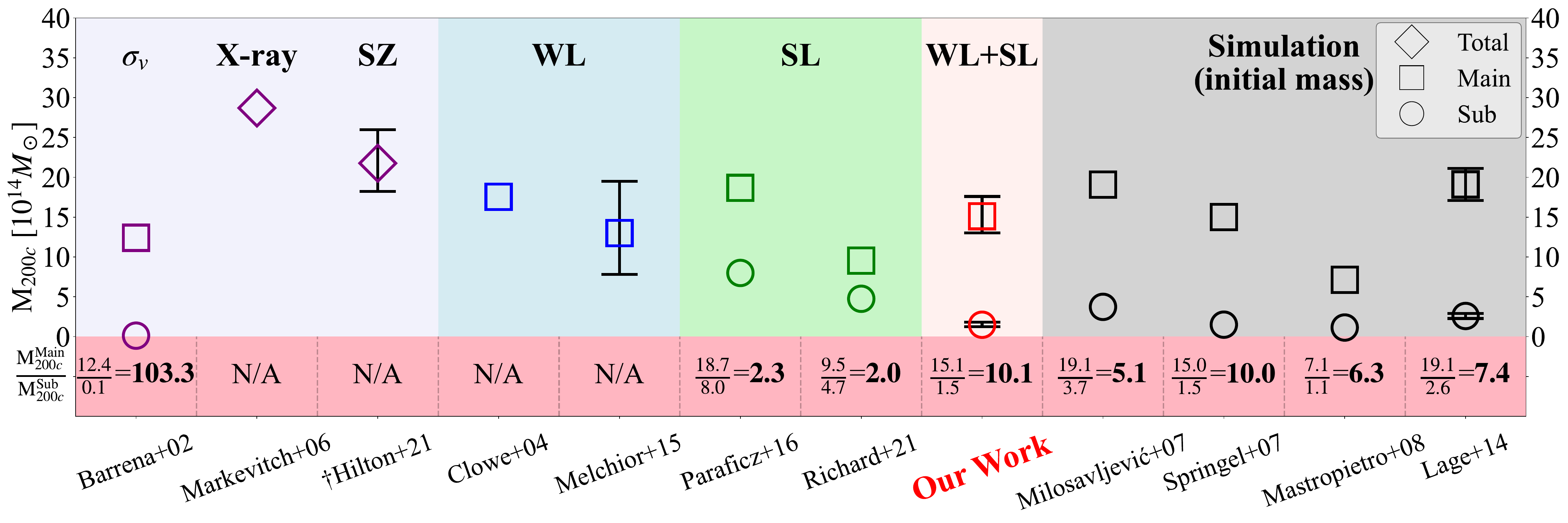} 
\caption{Compilation of Bullet Cluster virial mass ($M_{200c}$) measurements spanning more than two decades. Measurement methods are color-coded: equilibrium-based methods (purple: velocity dispersion, X-ray, SZ), WL (blue), SL (green), our joint WL+SL analysis (red), and hydrodynamical simulation initial conditions (gray). Symbols denote mass components: diamonds for total system, squares for main cluster, circles for subcluster. Error bars represent 68\% confidence intervals where provided in the original literature. Mass ratios ($M_{200c}^{\text{Main}}$/$M_{200c}^{\text{Sub}}$) appear below each study. `N/A' denotes cases where the result does not constrain a mass ratio.
Our joint analysis definitively establishes a minor-merger scenario (ratio $\mytilde$10:1), resolving the long-standing tension with simulation requirements. $\dagger$\citet{Hilton_2021} reported $M_{200m}$ which we present without conversion to $M_{200c}$.}
\label{fig:5_literature}
\end{figure*}

Figure~\ref{fig:5_literature} presents a comprehensive comparison of the Bullet Cluster virial mass ($M_{200c}$) measurements spanning more than two decades, revealing systematic discrepancies between methodological approaches. 
Among the three equilibrium-based methods (purple), only the velocity-dispersion study reported individual component masses, yielding an extreme mass ratio of $\mytilde$103:1.
None of the WL studies provided separate masses for the two components of the Bullet Cluster.
SL studies (green) did not quote virial masses explicitly, but when their published parametric models are extrapolated to large radii, they consistently imply major-merger scenarios with mass ratios of order $\mytilde$2:1.

The limitations of equilibrium-based methods are expected for such a violently merging system. \citet{Barrena_2002} obtained an implausibly high mass ratio of 103:1 using velocity dispersions, while \citet{Benavides_2023} estimated a subcluster-to-total system mass ratio of $0.07^{+0.16}_{-0.06}$ with a total system $M_{200c}$ of $(13\pm2)\times10^{14}M_{\odot}$. Unlike previous studies based on the dynamical equilibrium assumption, \citet{Foex_2017} included the caustic method, which does not rely on that assumption. They derived a total system $M_{200c}$ of $(11.5\pm2.8) \times 10^{14}M_{\odot}$ using the caustic method, while their equilibrium-based methods of virial and Jeans analyses yielded $\sim$$40\%$ higher $M_{200c}$ estimates of $(15.9^{+2.3}_{-2.1}) \times 10^{14}M_{\odot}$ and $(16.4^{+3.2}_{-2.7}) \times 10^{14}M_{\odot}$, respectively.

Although the X-ray analysis of \citet{Markevitch_2006} clearly resolved the two components in the gas distribution, the system’s extreme dynamical state rendered hydrostatic mass estimates unreliable. As a result, they provided only a total mass estimate for the entire system. Similarly, the SZ-based mass estimate from \citet{Hilton_2021} does not resolve the individual halos and provides only an integrated total mass.

While SL provides model-independent constraints without relying on equilibrium, it suffers from an equally fundamental limitation for determining virial masses, which depend on information (or assumptions) far beyond the SL-constrained core.
As an illustration, using the published dual Pseudo-Isothermal Elliptical profile parameters from \citet{Paraficz_2016} and \citet{Richard_2021}, we derived the implied virial masses. They differ nearly by a factor of two. This discrepancy arises entirely from their choice of truncation radius, a parameter unconstrained by SL data alone. \citet{Paraficz_2016} adopted $r_{\text{cut}} = 4413$~kpc ($=1000\arcsec$), while \citet{Richard_2021} used 1000~kpc. If we use the same $r_{\text{cut}}$, the results become nearly identical, demonstrating that these measurements are dictated by arbitrary parameter choices rather than observational constraints.

Previous WL studies faced different challenges. Ground-based observations with source densities of only $\mytilde$5--15~arcmin$^{-2}$ lacked the statistical power to constrain the masses of the individual components.
\citet{Clowe_2004} modeled the main cluster as a single halo rather than two components, obtaining virial parameters only for the main cluster while providing only a projected mass for the subcluster. \citet{Melchior_2015} fitted a single NFW profile to the entire system, similarly unable to separate the subcluster’s contribution. The HST analysis by \citet{Clowe_2006} achieved higher resolution but focused on two-dimensional mass reconstruction, without reporting virial mass estimates for the individual halos.

Hydrodynamical simulations of the Bullet Cluster consistently adopt minor-merger initial conditions in order to reproduce its observed morphology, shock properties, and dark matter--gas offsets. \citet{Milosavljevic_2007} employed a 5.1:1 mass ratio in gas-dynamical simulations and showed that such an encounter can generate the observed bow shock and the separation between the X-ray gas and dark matter peaks. \citet{Springel_2007} used a 10:1 mass ratio, demonstrating that a substantially more massive main cluster is required to reproduce the shock velocity ($\mytilde4700$~km~s$^{-1}$), mass-centroid offsets, and lensing features. \citet{Mastropietro_2008} adopted a 6.3:1 merger in high-resolution $N$-body/SPH simulations, finding good agreement with the displaced gas morphology and projected mass peaks.  \citet{Lage_2014} modeled the system with a 7.4:1 mass ratio in simulations that included triaxial halos and additional baryonic physics, likewise reproducing the observed shock structure and subcluster displacement. Collectively, these studies indicate that matching the Bullet Cluster’s key observational signatures requires a merger in which the main cluster is at least $\mytilde$5--10 times more massive than the subcluster, firmly supporting a minor-merger scenario.

\subsection{Robustness Tests with Alternative Configurations and Datasets}
\label{sec:model_comparison}

\begin{table*}
\caption{Mass Ratios and Model Comparison with Bayes Factors}
\centering
\setlength{\tabcolsep}{6pt}
\renewcommand{\arraystretch}{1.2}
\begin{tabular}{lcccclcccc}
\hline\hline
Config. & $M_{200c}^{\text{Main}}$ & $M_{200c}^{\text{Sub}}$ & Ratio & $\ln(\text{BF}_{cM})$ & 
Config. & $M_{200c}^{\text{Main}}$ & $M_{200c}^{\text{Sub}}$ & Ratio & $\ln(\text{BF}_{32})$ \\
\hline
$\bm{3cS}$ & $\bm{15.11^{+2.48}_{-2.10}}$ & $\bm{1.49^{+0.32}_{-0.25}}$ & $\bm{10.14^{+3.22}_{-2.47}}$ & \multirow{2}{*}{15.86} &
$3cL$ & $15.68^{+2.91}_{-2.46}$ & $1.54^{+0.33}_{-0.26}$ & $10.21^{+3.30}_{-2.57}$ & \multirow{2}{*}{38.30} \\
\cline{1-4}\cline{6-9}
$3MS$ & $7.36^{+0.67}_{-0.52}$ & $3.79^{+0.29}_{-0.29}$ & $1.95^{+0.26}_{-0.21}$ & &
$2cL$ & $19.18^{+3.27}_{-2.78}$ & $1.50^{+0.32}_{-0.26}$ & $12.77^{+4.00}_{-3.11}$ & \\
\hline
\end{tabular}
\tablecomments{Mass ratios and Bayes factors for different model configurations. All masses are in units of $10^{14}M_{\odot}$. Configuration notation follows Table~\ref{table:notation}. $M_{200c}^{\text{Main}}$ is measured at the midpoint between the two main BCGs. $M_{200c}^{\text{Sub}}$ corresponds to the subcluster component. The ratio column reports $M_{200c}^{\text{Main}} / M_{200c}^{\text{Sub}}$. {\tt 3MS} imposes the \citet{Diemer_2019} $M$--$c$ relation. The two-halo model {\tt 2cS} with SL constraints failed to converge because it cannot reproduce the observed SL-projected masses. The Bayes factors $\ln(\text{BF}_{cM})$ and $\ln(\text{BF}_{32})$ represent $\ln(\text{BF}_{(3cS)(3MS)})$ and $\ln(\text{BF}_{(3cL)(2cL)})$, respectively, where both {\tt 3MS} and {\tt 2cL} are strongly disfavored (log Bayes factor $>$ 15) compared to their corresponding reference configurations.}
\label{table:4_results_with_bayes}
\end{table*}

\begin{figure*}
\centering
\includegraphics[width=0.9\textwidth]{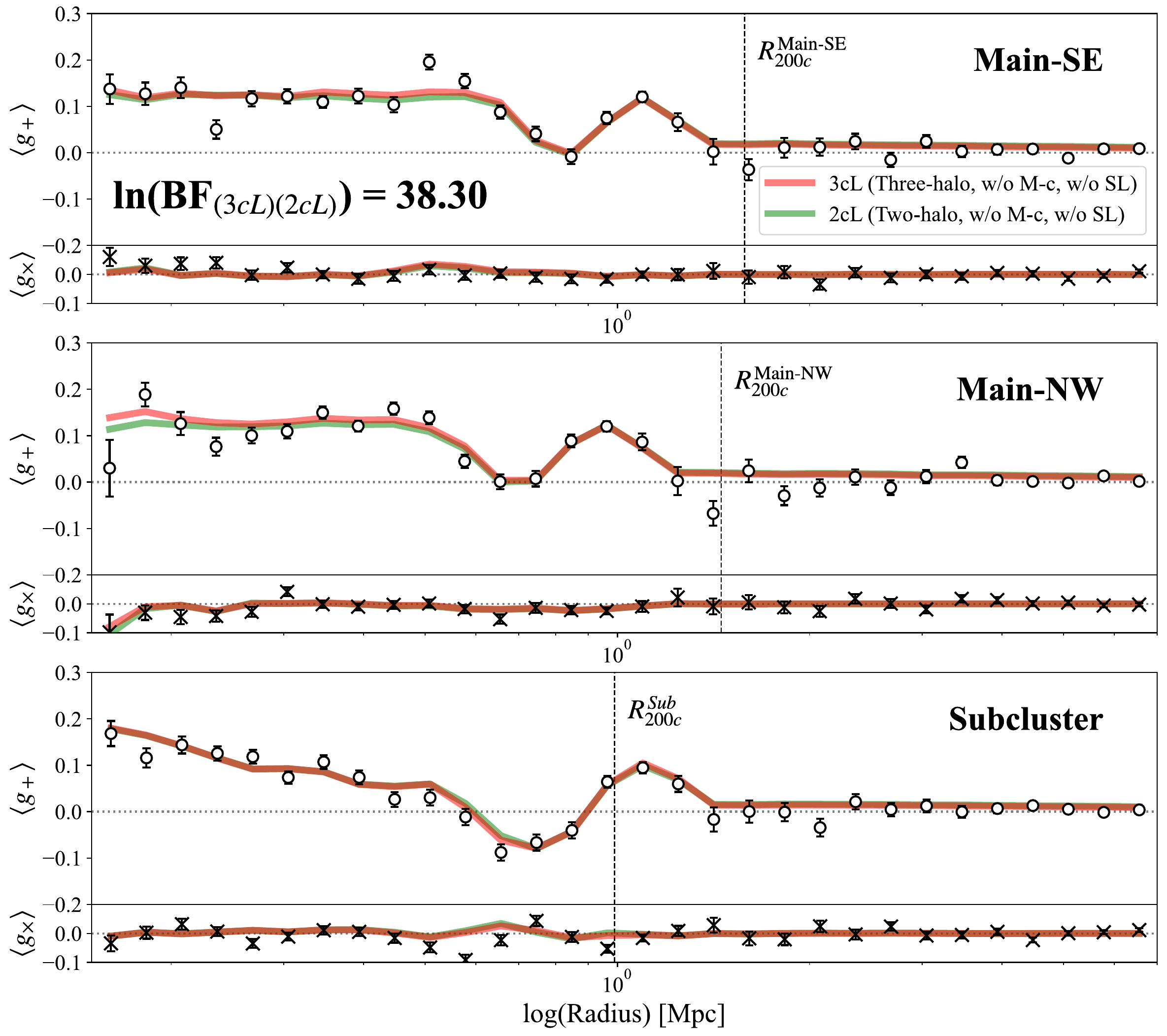} 
\caption{Tangential shear profile comparison for three-halo and two-halo NFW models. The three panels show measurements centered on Main-SE, Main-NW, and Sub BCGs, respectively, with tangential shear $\langle g_+\rangle$ (upper panels) and cross shear $\langle g_\times\rangle$ (lower panels). Black points and crosses with error bars represent observed shear measurements binned into 30 logarithmic radial annuli. Red lines show the {\tt 3cL} configuration, while green lines show the {\tt 2cL} configuration. Vertical dashed line indicates the virial radius of the reference halo. The overwhelming Bayes factor of $\ln(\text{BF}_{(3cL)(2cL)}) = 38.30$ provides decisive evidence for the three-halo model over the two-halo model, though both match the observed tangential shear profiles reasonably well.}
\label{fig:5_estim_tshear}
\end{figure*}

\subsubsection{Impact of Imposing an $M$--$c$ Relation}
To assess whether the $M$--$c$ relation is supported by the data, we first compare the fiducial model ({\tt 3cS}) with the configuration in which the \citet{Diemer_2019} $M$--$c$ relation is imposed ({\tt 3MS}), using the Bayes factor (Table~\ref{table:4_results_with_bayes}).
Employing bridge sampling to estimate the marginal likelihoods \citep{Gronau_2017}, we obtain
$\ln(\text{BF}_{(3cS)(3MS)}) = 15.86$, which constitutes decisive evidence in favor of the free-concentration (fiducial) model.
For robustness, we also compute the Bayes factor using three additional estimators: the harmonic mean \citep{Newton_1994}, stabilized harmonic mean \citep{Robert_2009}, and posterior mean \citep{Gelfand_1994}, which yield $\ln\mathrm{BF}=16.33$, $15.60$, and $15.57$, respectively. The consistency across all four methods confirms that the data strongly disfavor enforcing an $M$--$c$ relation for this system, consistent with the findings of \citet{Finner_2025} for highly disturbed clusters.

Given this statistical preference, it is instructive to examine how the imposed $M$--$c$ relation affects the parameter inference. Enforcing the relation drives the main-cluster concentration from $c_{200c}\sim2.0$ to $3.8$, which reduces its mass from $15.11^{+2.48}_{-2.10}\times10^{14}M_{\odot}$ to $7.36^{+0.67}_{-0.52}\times10^{14}M_{\odot}$. Conversely, the subcluster concentration is forced downward from $c_{200c}\sim9.7$ to $3.8$, increasing its mass from $1.49^{+0.32}_{-0.25}\times10^{14}M_{\odot}$ to $3.79^{+0.29}_{-0.29}\times10^{14}M_{\odot}$. These shifts change the inferred mass ratio from $\mytilde$10:1 to $\mytilde$2:1, implying a major-merger scenario that contradicts both our fiducial results and the requirements of hydrodynamical simulations.

\subsubsection{Limitations of the Two-Halo Model}
\label{sec:twohalofit}
To evaluate whether the main cluster can be represented as a single halo, we compare the two-halo configuration with the three-halo model. 
We find that the {\tt 2cS} configuration fails to produce a physically viable solution: a two-halo description cannot simultaneously reproduce the projected masses in the SL-anchoring regions and the wide-field WL shear measurements. As a result, the {\tt 2cS} model is ruled out at the level of parameter feasibility, and the {\tt 3cS} model is unambiguously preferred in this comparison.

Next, we compare the {\tt 3cL} and {\tt 2cL} configurations to test whether WL data alone can discriminate between the three-halo and two-halo models. Bridge sampling yields $\ln(\text{BF}_{(3cL)(2cL)}) = 38.30$, providing decisive evidence against the two-halo configuration. This strong model preference is enabled by the exceptional quality of the JWST WL data in the central region, which captures the distinct contributions of the two main-cluster halos. When we repeat the experiment using HST WL measurements in place of JWST, the discriminating power disappears, indicating that such model separation requires the depth and source density uniquely provided by JWST (\textsection\ref{sec:hst}).

Interestingly, we find that the tangential shears predicted from the {\tt 3cS}, {\tt 3cL}, and {\tt 2cL} models all match the observed measurements remarkably well.
Figure~\ref{fig:5_estim_tshear} presents the observed tangential shear profiles centered on the Main-SE (top), Main-NW (middle), and Subcluster (bottom) BCGs, together with the predictions from the {\tt 3cL} and {\tt 2cL} models. For clarity, we omit the prediction from the {\tt 3cS} (fiducial) model, as it is virtually indistinguishable from the {\tt 3cL} result. 
This agreement shows that the tangential shear alone does not provide sufficient discriminatory power to distinguish between the two- and three-halo models, as the detailed substructure information is largely washed out by azimuthal averaging.

\subsubsection{Testing Alternative Density Profiles: Einasto and Truncated NFW}
To evaluate whether alternative halo profiles offer any improvement over our fiducial NFW model, we test two widely used density parametrizations: the Einasto profile \citep{Einasto_1965} and the truncated NFW (TNFW; \citealt{Oguri_2011}) profile. Both models introduce additional flexibility beyond the standard NFW form: Einasto through a variable logarithmic slope and TNFW through an outer truncation radius. Given the complex dynamical state of the Bullet Cluster, it is important to assess whether these alternative descriptions provide a better fit to the WL+SL data or whether their additional freedom leads to parameter degeneracies and poorer physical performance.

The Einasto profile with the shape parameter fixed at $\alpha = 0.18$, the canonical value predicted for dark matter halos in $\Lambda$CDM \citep{Eckert_2022}, yields mass estimates that are broadly consistent with our fiducial NFW results: a main-cluster mass of $15.00^{+2.25}_{-2.04} \times 10^{14}M_{\odot}$, a subcluster mass of $1.50^{+0.33}_{-0.27} \times 10^{14}M_{\odot}$, and a mass ratio of $10.01^{+3.19}_{-2.48}$. However, the concentration of the Main-SE component reaches the lower prior boundary ($c_{200c}=1$), indicating that the fixed-$\alpha$ Einasto profile does not provide a better fit to the data than the standard NFW model.

Bridge sampling comparing our fiducial {\tt 3cS} NFW model with the {\tt 3cS} Einasto profile yields $\ln(\text{BF}) = 2.46$, with three additional estimators giving values in the range 2.58--2.82, indicating moderate evidence in favor of the simpler NFW model.
When $\alpha$ is allowed to vary freely, the chains fail to converge, suggesting that the added shape flexibility of the Einasto profile introduces non-negligible parameter degeneracies that cannot be constrained by the current WL+SL data in this dynamically complex merging system.

The TNFW profile, with truncation radius $r_t$ set by the prescription of \citet{Oguri_2011}, yields mass estimates that are also broadly consistent with our fiducial NFW results: a main-cluster mass of $15.35^{+3.74}_{-3.01}\times10^{14}M_{\odot}$, a subcluster mass of $1.45^{+0.43}_{-0.29}\times10^{14}M_{\odot}$, and a mass ratio of $9.88^{+4.17}_{-3.18}$.
As in the Einasto case, the concentration of the Main-SE component again approaches the lower prior boundary, indicating that the imposed truncation does not provide an improved description of the data.
Bridge sampling comparing the {\tt 3cS} NFW and {\tt 3cS} TNFW models yields $\ln(\text{BF}) = 13.28$,
with the three additional estimators spanning 3.75--17.78. The harmonic-mean value of 3.75 is known to be unstable, while the two robust estimators give consistent values of 17.77--17.78, providing decisive evidence in favor of the standard NFW profile over the truncated version.

This strong preference, much more pronounced than in the Einasto case, likely reflects the fact that the adopted truncation prescription may misrepresent the extreme dynamical state of the Bullet Cluster.
Whereas the Einasto profile modifies only the density slope, the TNFW model introduces a steep decline in the density profile beyond the truncation radius $r_t$; however, \citet{Walker_2025} show that high accretion rates in rapidly merging systems lead to much smaller $r_t$ values than those predicted by population-averaged prescriptions, making the standard truncation formula unsuitable. As with the Einasto profile, allowing $r_t$ to vary freely results in non-convergent chains, confirming that additional profile flexibility without informative priors cannot improve the fit for this highly disturbed system.

\begin{figure*}
\centering
\includegraphics[width=0.9\textwidth]{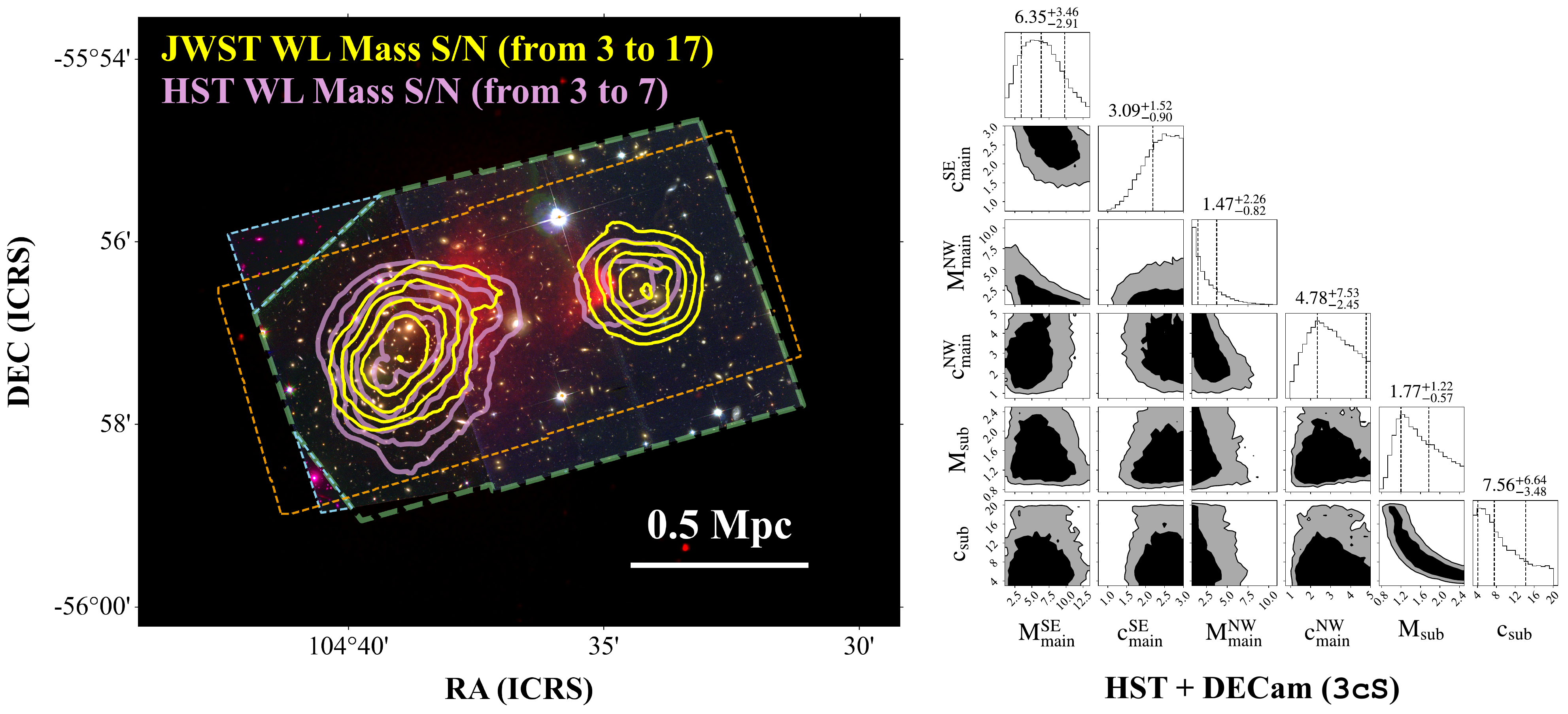}
\caption{Limitations of HST WL. \textit{Left}: HST-only mass reconstruction using {\tt FIATMAP} with S/N contours (purple, levels from 3 to 7 in steps of 1), achieving peak S/N of 7.90 (main cluster) and 4.63 (subcluster). JWST WL S/N contours are overlaid in yellow (levels from 3 to 17 in steps of 3.5), achieving peak values of 17.05 (main cluster) and 17.31 (subcluster). Shape measurements are performed with the F606W filter, and source selection yields a density of 87.8~arcmin$^{-2}$ using the F606W+F814W (green dashed region) and F606W+F850LP (blue dashed region) filter pairs. The orange rectangle marks the JWST coverage. Chandra X-ray emission (red) is overlaid onto HST-color image. \textit{Right}: Same as in Figure~\ref{fig:5_estim_corner} except that the analysis used the HST+DECam WL data. Although the HST analysis produces a mass map similar to JWST’s, it cannot robustly constrain three-halo models.
}
\label{fig:5_hst}
\end{figure*}

\subsubsection{Limitations of HST WL for Three-Halo Modeling}
\label{sec:hst}

We tested whether HST WL data, when combined with DECam, can also support three-halo modeling by performing a dedicated HST+DECam analysis. We retrieved the Advanced Camera for Surveys (ACS) images from the STScI archive. WL analysis was performed with the F606W filter. 
We constructed an extensive PSF library from 342 stellar-field exposures taken over nearly two decades (2002-04-17 to 2021-07-20). 
For each Bullet Cluster exposure, we selected the optimal PSF template, modeled the PSF on the individual frames, and stacked these to generate a final mosaic PSF model. 
We based our source selection on the F606W-F814W and F606W-F850LP colors.
The resulting WL source density is 87.8~arcmin$^{-2}$. Readers are referred to our previous papers \cite[e.g.,][]{Jee_2006,Jee_2009,Kim_2021} for further details on our ACS WL pipeline.
Figure~\ref{fig:5_hst} (left) shows the HST WL mass reconstruction produced with {\tt FIATMAP}, with JWST WL S/N contours overlaid for direct comparison.

This comparatively low source density limits the constraining power of HST WL and prevents stable three-halo modeling. As shown in the right panel of Figure~\ref{fig:5_hst}, the HST+DECam three-halo fits exhibit strong parameter degeneracies, yielding broad and weakly constrained posteriors even when SL anchoring and wide-field DECam coverage are included. The resulting HST+DECam {\tt 3cS} masses $M_{200c}^{\mathrm{Main}} = 10.03^{+3.18}_{-2.59}\times10^{14}M_{\odot}$ and $M_{200c}^{\mathrm{Sub}} = 1.77^{+1.22}_{-0.57}\times10^{14}M_{\odot}$ imply a mass ratio of $5.58^{+4.19}_{-2.70}$ and a total system mass of $13.66^{+3.42}_{-2.87}\times10^{14}M_{\odot}$, all substantially offset from our fiducial JWST+DECam results.

These large discrepancies demonstrate that, while HST provides high-quality WL data for many cluster studies, its effective source density in this field is insufficient to resolve the distinct components of the Bullet Cluster. JWST WL measurements are therefore essential for robust three-halo modeling.

\subsubsection{Need for Wide-Field DECam Coverage}
\label{sec:jwstonly}

\begin{figure*}
\centering
\includegraphics[width=0.9\textwidth]{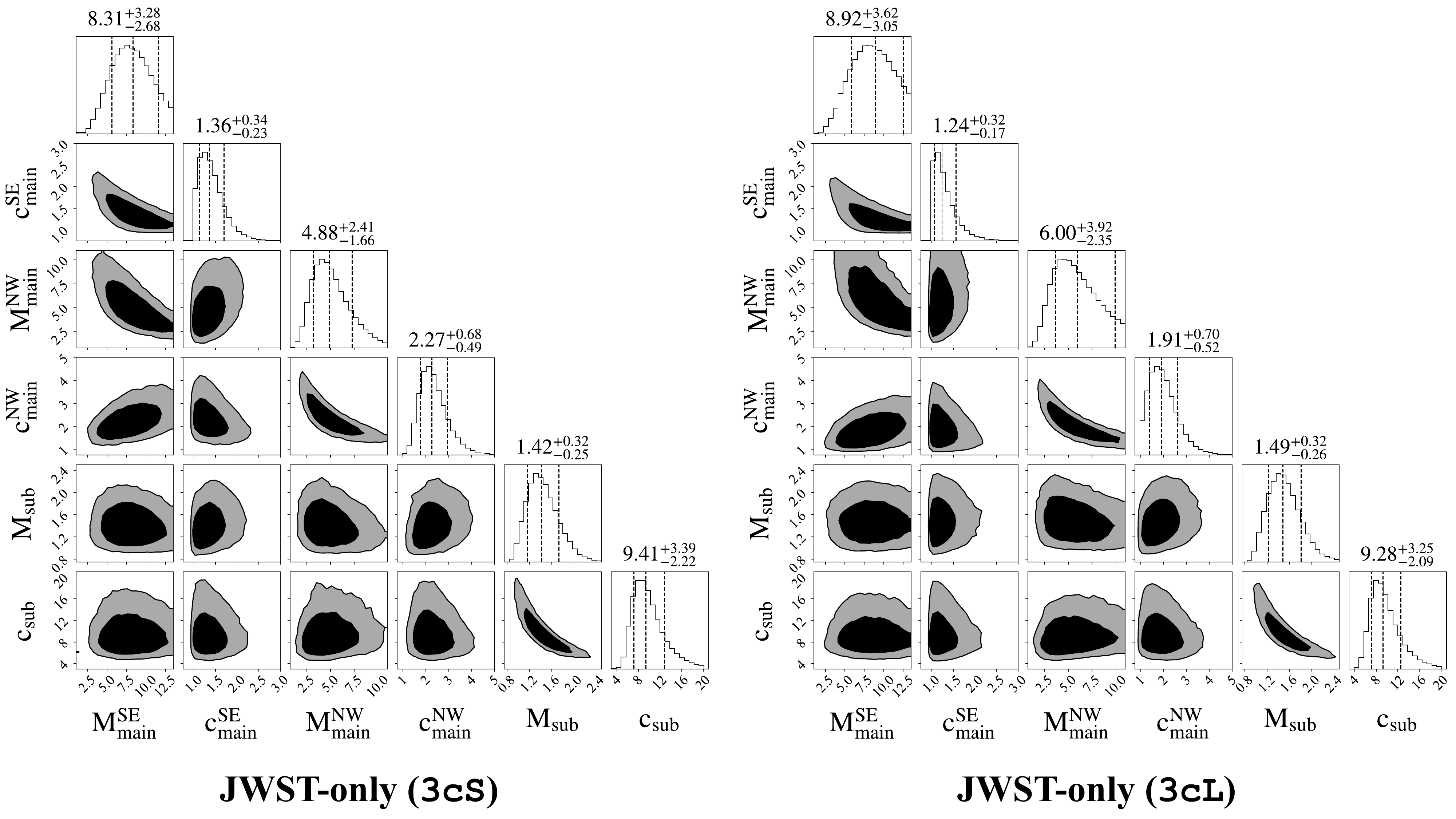} 
\caption{Same as Figure~\ref{fig:5_estim_corner} except that the results are for JWST-only {\tt 3cS} and {\tt 3cL} configurations. }
\label{fig:5_jwstonly_corner}
\end{figure*}

Although JWST provides an extraordinary source density of $\mytilde400$~arcmin$^{-2}$, mass estimation without DECam WL coverage leads to a non-negligible overestimate due to forced extrapolation beyond the JWST field. In the JWST-only {\tt 3cS} configuration, the inferred total system's mass increases by $\mytilde$21\% relative to the fiducial JWST+DECam result, while the {\tt 3cL} configuration produces an even larger bias of $\mytilde$35\%.

The posterior distributions also show signs of instability: Main-SE’s concentration parameter collapses toward the lower prior boundary at $c_{200c}=1$ (Figure~\ref{fig:5_jwstonly_corner}), and Main-NW exhibits similar, though weaker, boundary behavior. Incorporating SL constraints (JWST-only {\tt 3cS}) alleviates these effects, stabilizing Main-NW and partially stabilizing Main-SE, while reducing the total system's mass overestimate from 35\% to 21\%. This demonstrates that SL provides valuable central normalization but cannot by itself compensate for the absence of wide-field WL information.
In contrast, the subcluster’s high concentration and low mass place its shear signal mostly within the JWST field, allowing its mass and concentration to remain consistent with fiducial values.

\subsubsection{Impact of Including SL Constraints}
We compare our fiducial joint WL+SL model ({\tt 3cS}) with the WL-only configuration ({\tt 3cL}) in
Table~\ref{table:4_main_results}. The two sets of results are statistically consistent, although as expected, the inclusion of SL anchoring in {\tt 3cS} reduces the width of the posterior distributions. This implies that the projected masses within the anchoring regions are in good agreement with the {\tt 3cL} WL model extrapolated inward.

At face value, the close agreement between the two models may appear at odds with the general conclusion of \citet{Lee_2023}, who showed that WL-only mass inference in merging clusters can suffer from substantial model bias. However, their simulations demonstrate that the magnitude of this bias strongly depends on merger mass ratio: it is largest for major mergers and gradually decreases as the system approaches the minor-merger regime. For a $\mytilde$10:1 merger, \citet{Lee_2023} predict biases of $\mytilde$20\%, substantially smaller than those for major-merger configurations. Given the statistical uncertainties in our measurements, the level of agreement between {\tt 3cS} and {\tt 3cL} is therefore expected rather than surprising.
If the Bullet Cluster had been a major merger, the difference between the {\tt 3cS} and {\tt 3cL} results would likely have been substantial.

\section{Conclusion}
\label{sec:con}
We have presented the first robust virial mass determination of the Bullet Cluster derived from a joint weak- and strong-lensing analysis that combines high-resolution JWST/NIRCam data with wide-field DECam imaging. This integrated approach resolves long-standing uncertainties in the system’s individual masses, mass ratio, and merger configuration that have persisted for more than two decades.

Our analysis yields $M_{200c} = 15.11^{+2.48}_{-2.10}\times10^{14}M_{\odot}$ for the main cluster and $1.49^{+0.32}_{-0.25}\times10^{14}M_{\odot}$ for the subcluster, corresponding to a mass ratio of $10.14^{+3.22}_{-2.47}$. This result provides the first observational confirmation that the Bullet Cluster is a $\mytilde$10:1 minor merger, fully consistent with the mass ratios required by hydrodynamical simulations to reproduce its shock properties, gas--dark matter offsets, and overall morphology. The mass ratio remains stable across a wide range of systematic tests.

Three factors enable this improved precision.
(1) JWST’s exceptional background source density ($\mytilde400$~arcmin$^{-2}$) resolves three distinct halos in the cluster core, allowing independent characterization of the two main-cluster components and the subcluster.
(2) DECam provides shear constraints to radii $\mytilde3R_{200c}^{\text{Total}}$, eliminating the need for extrapolation that dominated the uncertainty budget in earlier SL-only studies.
(3) Model-independent projected masses from JWST strong lensing serve as anchoring conditions for NFW profile fitting, reducing the mass--concentration degeneracy and mitigating model biases intrinsic to merging systems.

Beyond resolving the virial mass ratio, our three-halo reconstruction reveals that the main cluster is itself bimodal: two comparable-mass halos (ratio $\mytilde$1.3:1) with low concentrations and a projected separation of 170~kpc. The axis connecting them is tilted by $\mytilde50^\circ$ relative to the bullet’s primary E-W collision axis, suggesting that the main cluster experienced a prior merger. This three-body configuration supersedes the simplified two-halo initial conditions commonly adopted in Bullet Cluster simulations and provides a more realistic framework for future hydrodynamical modeling.

The methodology developed here, leveraging space-based resolution, wide-field coverage, and WL+SL anchoring, establishes a general pathway for accurate mass calibration of dynamically complex clusters. As Euclid and the Nancy Grace Roman Space Telescope deliver deep, wide-area imaging for thousands of clusters, similar analyses will enable substantially improved constraints on cluster mass functions, scaling relations, and ultimately the properties of dark matter and dark energy.

\begin{acknowledgments}
We are grateful to Wonki Lee, Hyosun Park, Eunmo Ahn, Kyle Finner, and Zachary Scofield for providing useful comments. We acknowledge Frank Valdes for his assistance with the initial DECam data reduction. Based on observations made with the NASA/ESA/CSA James Webb Space Telescope (JWST) and NASA/ESA Hubble Space Telescope (HST), the data were obtained from the Mikulski Archive for Space Telescopes (MAST) at the Space Telescope Science Institute (STScI). These space-based observations analyzed can be accessed via \dataset[doi:10.17909/8zea-jv19]{https://doi.org/10.17909/8zea-jv19}. STScI is operated by the Association of Universities for Research in Astronomy, Inc., under NASA contract NAS 5-03127. MJJ acknowledges support for the current research from the National Research Foundation (NRF) of Korea under the programs 2022R1A2C1003130 and RS-2023-00219959.
\end{acknowledgments}

\bibliographystyle{aasjournal}
\bibliography{main}

@article{Bradac_2006,
  author       = {Brada{\v c}, Maru{\v s}a and Clowe, Douglas and Gonzalez, Anthony H. and Marshall, Phil and Forman, William and Jones, Christine and Markevitch, Maxim and Randress, H{\aa}kon},
  title        = {Strong and Weak Lensing United: The Total Mass Distribution in the Merging Cluster 1E 0657-56},
  journal      = {The Astrophysical Journal},
  volume       = {652},
  number       = {2},
  pages        = {937--947},
  year         = {2006},
  doi          = {10.1086/508601}
}

@ARTICLE{Jee_2006,
       author = {{Jee}, M.~J. and {White}, R.~L. and {Ford}, H.~C. and {Illingworth}, G.~D. and {Blakeslee}, J.~P. and {Holden}, B. and {Mei}, S.},
        title = "{Weak-lensing Detection at z \raisebox{-0.5ex}\textasciitilde 1.3: Measurement of the Two Lynx Clusters with the Advanced Camera for Surveys}",
      journal = {\apj},
         year = 2006,
        month = may,
       volume = {642},
       number = {2},
        pages = {720-733},
          doi = {10.1086/501427},
archivePrefix = {arXiv},
       eprint = {astro-ph/0601334},
 primaryClass = {astro-ph},
       adsurl = {https://ui.adsabs.harvard.edu/abs/2006ApJ...642..720J}
}

@ARTICLE{Kim_2021,
       author = {{Kim}, Jinhyub and {Jee}, M. James and {Hughes}, John P. and {Yoon}, Mijin and {HyeongHan}, Kim and {Menanteau}, Felipe and {Sif{\'o}n}, Crist{\'o}bal and {Hovey}, Luke and {Arunachalam}, Prasiddha},
        title = "{Head-to-Toe Measurement of El Gordo: Improved Analysis of the Galaxy Cluster ACT-CL J0102-4915 with New Wide-field Hubble Space Telescope Imaging Data}",
      journal = {\apj},
         year = 2021,
        month = dec,
       volume = {923},
       number = {1},
          eid = {101},
        pages = {101},
          doi = {10.3847/1538-4357/ac294f},
archivePrefix = {arXiv},
       eprint = {2106.00031},
 primaryClass = {astro-ph.CO},
       adsurl = {https://ui.adsabs.harvard.edu/abs/2021ApJ...923..101K}
}

@ARTICLE{Jee_2009,
       author = {{Jee}, M. James and {Tyson}, J. Anthony},
        title = "{Dark Matter in the Galaxy Cluster CL J1226+3332 at z = 0.89}",
      journal = {\apj},
         year = 2009,
        month = feb,
       volume = {691},
       number = {2},
        pages = {1337-1347},
          doi = {10.1088/0004-637X/691/2/1337},
archivePrefix = {arXiv},
       eprint = {0810.0709},
 primaryClass = {astro-ph},
       adsurl = {https://ui.adsabs.harvard.edu/abs/2009ApJ...691.1337J}
}

@article{Finner_2023,
doi = {10.3847/1538-4357/ace1e6},
url = {https://dx.doi.org/10.3847/1538-4357/ace1e6},
year = {2023},
month = {aug},
publisher = {The American Astronomical Society},
volume = {953},
number = {1},
pages = {102},
author = {Finner, Kyle and Faisst, Andreas and Chary, Ranga-Ram and Jee, M. James},
title = {The First Weak-lensing Analysis with the James Webb Space Telescope: SMACS J0723.3–7327},
journal = {The Astrophysical Journal}
}

@article{Cha_2025,
doi = {10.3847/2041-8213/add2f0},
url = {https://dx.doi.org/10.3847/2041-8213/add2f0},
year = {2025},
month = {jun},
publisher = {The American Astronomical Society},
volume = {987},
number = {1},
pages = {L15},
author = {Cha, Sangjun and Cho, Boseong Young and Joo, Hyungjin and Lee, Wonki and HyeongHan, Kim and Scofield, Zachary P. and Finner, Kyle and Jee, M. James},
title = {A High-Caliber View of the Bullet Cluster through JWST Strong and Weak Lensing Analyses},
journal = {The Astrophysical Journal Letters}
}

@software{Bertin_2010,
       author = {{Bertin}, Emmanuel},
        title = "{SWarp: Resampling and Co-adding FITS Images Together}",
 howpublished = {Astrophysics Source Code Library, record ascl:1010.068},
         year = 2010,
        month = oct,
          eid = {ascl:1010.068},
       adsurl = {https://ui.adsabs.harvard.edu/abs/2010ascl.soft10068B}
}

@INPROCEEDINGS{Bertin_2006,
       author = {{Bertin}, E.},
        title = "{Automatic Astrometric and Photometric Calibration with SCAMP}",
    booktitle = {Astronomical Data Analysis Software and Systems XV},
         year = 2006,
       editor = {{Gabriel}, C. and {Arviset}, C. and {Ponz}, D. and {Enrique}, S.},
       series = {Astronomical Society of the Pacific Conference Series},
       volume = {351},
        month = jul,
        pages = {112},
       adsurl = {https://ui.adsabs.harvard.edu/abs/2006ASPC..351..112B}
}

@article{Gruen_2014,
doi = {10.1086/675080},
url = {https://dx.doi.org/10.1086/675080},
year = {2014},
month = {feb},
publisher = {University of Chicago Press},
volume = {126},
number = {936},
pages = {158},
author = {Gruen, D. and Seitz, S. and Bernstein, G. M.},
title = {Implementation of Robust Image Artifact Removal in SWarp through Clipped Mean Stacking},
journal = {Publications of the Astronomical Society of the Pacific}
}

@ARTICLE{Bertin_1996,
       author = {{Bertin}, E. and {Arnouts}, S.},
        title = "{SExtractor: Software for source extraction.}",
      journal = {\aaps},
         year = 1996,
        month = jun,
       volume = {117},
        pages = {393-404},
          doi = {10.1051/aas:1996164},
       adsurl = {https://ui.adsabs.harvard.edu/abs/1996A&AS..117..393B}
}

@article{Foex_2017,
	author = {{Foëx} and {Böhringer} and {Chon}},
	title = {Comparison of hydrostatic and dynamical masses of distant X-ray luminous galaxy clusters⋆⋆⋆},
	DOI= "10.1051/0004-6361/201731104",
	url= "https://doi.org/10.1051/0004-6361/201731104",
	journal = {A\&A},
	year = 2017,
	volume = 606,
	pages = "A122",
}

@article{Puccetti_2020,
	author = {{Puccetti} and {Fiore} and {Bongiorno} and {Boutsia} and {Fassbender} and {Verdugo}},
	title = {Triggering nuclear and galaxy activity in the Bullet cluster⋆},
	DOI= "10.1051/0004-6361/201833601",
	url= "https://doi.org/10.1051/0004-6361/201833601",
	journal = {A\&A},
	year = 2020,
	volume = 634,
	pages = "A137",
}

@article{Barrena_2002,
	author = {{Barrena} and {Biviano} and {Ramella} and {Falco} and {Seitz}},
	title = {The dynamical status of the cluster of galaxies 1E0657-56*},
	DOI= "10.1051/0004-6361:20020244",
	url= "https://doi.org/10.1051/0004-6361:20020244",
	journal = {A\&A},
	year = 2002,
	volume = 386,
	number = 3,
	pages = "816-828",
}

@article{Finner_2017,
doi = {10.3847/1538-4357/aa998c},
url = {https://dx.doi.org/10.3847/1538-4357/aa998c},
year = {2017},
month = {dec},
publisher = {The American Astronomical Society},
volume = {851},
number = {1},
pages = {46},
author = {Finner, Kyle and Jee, M. James and Golovich, Nathan and Wittman, David and Dawson, William and Gruen, Daniel and Koekemoer, Anton M. and Lemaux, Brian C. and Seitz, Stella},
title = {MC2: Subaru and Hubble Space Telescope Weak-lensing Analysis of the Double Radio Relic Galaxy Cluster PLCK G287.0+32.9},
journal = {The Astrophysical Journal}
}

@ARTICLE{Monet_2003,
       author = {{Monet}, David G. and {Levine}, Stephen E. and {Canzian}, Blaise and {Ables}, Harold D. and {Bird}, Alan R. and {Dahn}, Conard C. and {Guetter}, Harry H. and {Harris}, Hugh C. and {Henden}, Arne A. and {Leggett}, Sandy K. and {Levison}, Harold F. and {Luginbuhl}, Christian B. and {Martini}, Joan and {Monet}, Alice K.~B. and {Munn}, Jeffrey A. and {Pier}, Jeffrey R. and {Rhodes}, Albert R. and {Riepe}, Betty and {Sell}, Stephen and {Stone}, Ronald C. and {Vrba}, Frederick J. and {Walker}, Richard L. and {Westerhout}, Gart and {Brucato}, Robert J. and {Reid}, I. Neill and {Schoening}, William and {Hartley}, M. and {Read}, M.~A. and {Tritton}, S.~B.},
        title = "{The USNO-B Catalog}",
      journal = {\aj},
         year = 2003,
        month = feb,
       volume = {125},
       number = {2},
        pages = {984-993},
          doi = {10.1086/345888},
archivePrefix = {arXiv},
       eprint = {astro-ph/0210694},
 primaryClass = {astro-ph},
       adsurl = {https://ui.adsabs.harvard.edu/abs/2003AJ....125..984M}
}

@ARTICLE{Jester_2005,
       author = {{Jester}, Sebastian and {Schneider}, Donald P. and {Richards}, Gordon T. and {Green}, Richard F. and {Schmidt}, Maarten and {Hall}, Patrick B. and {Strauss}, Michael A. and {Vanden Berk}, Daniel E. and {Stoughton}, Chris and {Gunn}, James E. and {Brinkmann}, Jon and {Kent}, Stephen M. and {Smith}, J. Allyn and {Tucker}, Douglas L. and {Yanny}, Brian},
        title = "{The Sloan Digital Sky Survey View of the Palomar-Green Bright Quasar Survey}",
      journal = {\aj},
         year = 2005,
        month = sep,
       volume = {130},
       number = {3},
        pages = {873-895},
          doi = {10.1086/432466},
archivePrefix = {arXiv},
       eprint = {astro-ph/0506022},
 primaryClass = {astro-ph},
       adsurl = {https://ui.adsabs.harvard.edu/abs/2005AJ....130..873J}
}

@article{Abbott_2018,
doi = {10.3847/1538-4365/aae9f0},
url = {https://dx.doi.org/10.3847/1538-4365/aae9f0},
year = {2018},
month = {nov},
publisher = {The American Astronomical Society},
volume = {239},
number = {2},
pages = {18},
author = {Abbott, T. M. C. and Abdalla, F. B. and Allam, S. and Amara, A. and Annis, J. and Asorey, J. and Avila, S. and Ballester, O. and Banerji, M. and Barkhouse, W. and Baruah, L. and Baumer, M. and Bechtol, K. and Becker, M. R. and Benoit-Lévy, A. and Bernstein, G. M. and Bertin, E. and Blazek, J. and Bocquet, S. and Brooks, D. and Brout, D. and Buckley-Geer, E. and Burke, D. L. and Busti, V. and Campisano, R. and Cardiel-Sas, L. and Rosell, A. Carnero and Kind, M. Carrasco and Carretero, J. and Castander, F. J. and Cawthon, R. and Chang, C. and Chen, X. and Conselice, C. and Costa, G. and Crocce, M. and Cunha, C. E. and D’Andrea, C. B. and Costa, L. N. da and Das, R. and Daues, G. and Davis, T. M. and Davis, C. and Vicente, J. De and DePoy, D. L. and DeRose, J. and Desai, S. and Diehl, H. T. and Dietrich, J. P. and Dodelson, S. and Doel, P. and Drlica-Wagner, A. and Eifler, T. F. and Elliott, A. E. and Evrard, A. E. and Farahi, A. and Neto, A. Fausti and Fernandez, E. and Finley, D. A. and Flaugher, B. and Foley, R. J. and Fosalba, P. and Friedel, D. N. and Frieman, J. and García-Bellido, J. and Gaztanaga, E. and Gerdes, D. W. and Giannantonio, T. and Gill, M. S. S. and Glazebrook, K. and Goldstein, D. A. and Gower, M. and Gruen, D. and Gruendl, R. A. and Gschwend, J. and Gupta, R. R. and Gutierrez, G. and Hamilton, S. and Hartley, W. G. and Hinton, S. R. and Hislop, J. M. and Hollowood, D. and Honscheid, K. and Hoyle, B. and Huterer, D. and Jain, B. and James, D. J. and Jeltema, T. and Johnson, M. W. G. and Johnson, M. D. and Kacprzak, T. and Kent, S. and Khullar, G. and Klein, M. and Kovacs, A. and Koziol, A. M. G. and Krause, E. and Kremin, A. and Kron, R. and Kuehn, K. and Kuhlmann, S. and Kuropatkin, N. and Lahav, O. and Lasker, J. and Li, T. S. and Li, R. T. and Liddle, A. R. and Lima, M. and Lin, H. and López-Reyes, P. and MacCrann, N. and Maia, M. A. G. and Maloney, J. D. and Manera, M. and March, M. and Marriner, J. and Marshall, J. L. and Martini, P. and McClintock, T. and McKay, T. and McMahon, R. G. and Melchior, P. and Menanteau, F. and Miller, C. J. and Miquel, R. and Mohr, J. J. and Morganson, E. and Mould, J. and Neilsen, E. and Nichol, R. C. and Nogueira, F. and Nord, B. and Nugent, P. and Nunes, L. and Ogando, R. L. C. and Old, L. and Pace, A. B. and Palmese, A. and Paz-Chinchón, F. and Peiris, H. V. and Percival, W. J. and Petravick, D. and Plazas, A. A. and Poh, J. and Pond, C. and Porredon, A. and Pujol, A. and Refregier, A. and Reil, K. and Ricker, P. M. and Rollins, R. P. and Romer, A. K. and Roodman, A. and Rooney, P. and Ross, A. J. and Rykoff, E. S. and Sako, M. and Sanchez, M. L. and Sanchez, E. and Santiago, B. and Saro, A. and Scarpine, V. and Scolnic, D. and Serrano, S. and Sevilla-Noarbe, I. and Sheldon, E. and Shipp, N. and Silveira, M. L. and Smith, M. and Smith, R. C. and Smith, J. A. and Soares-Santos, M. and Sobreira, F. and Song, J. and Stebbins, A. and Suchyta, E. and Sullivan, M. and Swanson, M. E. C. and Tarle, G. and Thaler, J. and Thomas, D. and Thomas, R. C. and Troxel, M. A. and Tucker, D. L. and Vikram, V. and Vivas, A. K. and Walker, A. R. and Wechsler, R. H. and Weller, J. and Wester, W. and Wolf, R. C. and Wu, H. and Yanny, B. and Zenteno, A. and Zhang, Y. and Zuntz, J. and (DES Collaboration) and Juneau, S. and Fitzpatrick, M. and Nikutta, R. and Nidever, D. and Olsen, K. and Scott, A. and (NOAO Data Lab)},
title = {The Dark Energy Survey: Data Release 1},
journal = {The Astrophysical Journal Supplement Series}
}

@ARTICLE{Wright_2000,
       author = {{Wright}, Candace Oaxaca and {Brainerd}, Tereasa G.},
        title = "{Gravitational Lensing by NFW Halos}",
      journal = {\apj},
         year = 2000,
        month = may,
       volume = {534},
       number = {1},
        pages = {34-40},
          doi = {10.1086/308744},
       adsurl = {https://ui.adsabs.harvard.edu/abs/2000ApJ...534...34W}
}

@article{Bartelmann_2001,
title = {Weak gravitational lensing},
journal = {Physics Reports},
volume = {340},
number = {4},
pages = {291-472},
year = {2001},
issn = {0370-1573},
doi = {https://doi.org/10.1016/S0370-1573(00)00082-X},
url = {https://www.sciencedirect.com/science/article/pii/S037015730000082X},
author = {Matthias Bartelmann and Peter Schneider}
}

@ARTICLE{Schneider_2005,
       author = {{Schneider}, Peter},
        title = "{Weak Gravitational Lensing}",
      journal = {arXiv e-prints},
         year = 2005,
        month = sep,
          eid = {astro-ph/0509252},
        pages = {astro-ph/0509252},
          doi = {10.48550/arXiv.astro-ph/0509252},
archivePrefix = {arXiv},
       eprint = {astro-ph/0509252},
 primaryClass = {astro-ph},
       adsurl = {https://ui.adsabs.harvard.edu/abs/2005astro.ph..9252S}
}

@Article{Hoekstra_2013,
author={Hoekstra, Henk
and Bartelmann, Matthias
and Dahle, H{\aa}kon
and Israel, Holger
and Limousin, Marceau
and Meneghetti, Massimo},
title={Masses of Galaxy Clusters from Gravitational Lensing},
journal={Space Science Reviews},
year={2013},
month={Aug},
day={01},
volume={177},
number={1},
pages={75-118},
issn={1572-9672},
doi={10.1007/s11214-013-9978-5},
url={https://doi.org/10.1007/s11214-013-9978-5}
}

@article{Mandelbaum_2018,
   author = "Mandelbaum, Rachel",
   title = "Weak Lensing for Precision Cosmology", 
   journal= "Annual Review of Astronomy and Astrophysics",
   year = "2018",
   volume = "56",
   number = "Volume 56, 2018",
   pages = "393-433",
   doi = "https://doi.org/10.1146/annurev-astro-081817-051928",
   url = "https://www.annualreviews.org/content/journals/10.1146/annurev-astro-081817-051928",
   publisher = "Annual Reviews",
   issn = "1545-4282",
   type = "Journal Article",
  }

@ARTICLE{Hoekstra_2008,
       author = {{Hoekstra}, Henk and {Jain}, Bhuvnesh},
        title = "{Weak Gravitational Lensing and Its Cosmological Applications}",
      journal = {Annual Review of Nuclear and Particle Science},
         year = 2008,
        month = nov,
       volume = {58},
       number = {1},
        pages = {99-123},
          doi = {10.1146/annurev.nucl.58.110707.171151},
archivePrefix = {arXiv},
       eprint = {0805.0139},
 primaryClass = {astro-ph},
       adsurl = {https://ui.adsabs.harvard.edu/abs/2008ARNPS..58...99H}
}

@article{Kilbinger_2015,
doi = {10.1088/0034-4885/78/8/086901},
url = {https://dx.doi.org/10.1088/0034-4885/78/8/086901},
year = {2015},
month = {jul},
publisher = {IOP Publishing},
volume = {78},
number = {8},
pages = {086901},
author = {Kilbinger, Martin},
title = {Cosmology with cosmic shear observations: a review},
journal = {Reports on Progress in Physics}
}

@INPROCEEDINGS{Markwardt_2009,
       author = {{Markwardt}, C.~B.},
        title = "{Non-linear Least-squares Fitting in IDL with MPFIT}",
    booktitle = {Astronomical Data Analysis Software and Systems XVIII},
         year = 2009,
       editor = {{Bohlender}, D.~A. and {Durand}, D. and {Dowler}, P.},
       series = {Astronomical Society of the Pacific Conference Series},
       volume = {411},
        month = sep,
        pages = {251},
          doi = {10.48550/arXiv.0902.2850},
archivePrefix = {arXiv},
       eprint = {0902.2850},
 primaryClass = {astro-ph.IM},
       adsurl = {https://ui.adsabs.harvard.edu/abs/2009ASPC..411..251M}
}

@ARTICLE{Ahn_2024,
       author = {{Ahn}, Eunmo and {Jee}, M. James and {Lee}, Wonki and {Joo}, Hyungjin and {ZuHone}, John},
        title = "{Substructures within Substructures in the Complex Postmerging System A514 Unveiled by High-resolution Magellan/Megacam Weak Lensing}",
      journal = {\apj},
         year = 2024,
        month = oct,
       volume = {973},
       number = {2},
          eid = {79},
        pages = {79},
          doi = {10.3847/1538-4357/ad65dc},
archivePrefix = {arXiv},
       eprint = {2404.04321},
 primaryClass = {astro-ph.CO},
       adsurl = {https://ui.adsabs.harvard.edu/abs/2024ApJ...973...79A}
}

@article{HyeongHan_2020,
doi = {10.3847/1538-4357/aba742},
url = {https://dx.doi.org/10.3847/1538-4357/aba742},
year = {2020},
month = {sep},
publisher = {The American Astronomical Society},
volume = {900},
number = {2},
pages = {127},
author = {Kim HyeongHan and M. James Jee and Lawrence Rudnick and David Parkinson and Kyle Finner and Mijin Yoon and Wonki Lee and Gianfranco Brunetti and Marcus Brüggen and Jordan D. Collier and Andrew M. Hopkins and Michał J. Michałowski and Ray P. Norris and Chris Riseley},
title = {Discovery of a Radio Relic in the Massive Merging Cluster SPT-CL J2023-5535 from the ASKAP-EMU Pilot Survey},
journal = {The Astrophysical Journal}
}

@article{Jee_2007,
doi = {10.1086/524849},
url = {https://dx.doi.org/10.1086/524849},
year = {2007},
month = {dec},
publisher = {University of Chicago Press},
volume = {119},
number = {862},
pages = {1403},
author = {M. J. Jee and J. P. Blakeslee and M. Sirianni and A. R. Martel and R. L. White and H. C. Ford},
title = {Principal Component Analysis of the Time- and Position-dependent Point-Spread Function of the Advanced Camera for Surveys},
journal = {Publications of the Astronomical Society of the Pacific}
}

@article{Jee_2013,
doi = {10.1088/0004-637X/765/1/74},
url = {https://dx.doi.org/10.1088/0004-637X/765/1/74},
year = {2013},
month = {feb},
publisher = {The American Astronomical Society},
volume = {765},
number = {1},
pages = {74},
author = {M. James Jee and J. Anthony Tyson and Michael D. Schneider and David Wittman and Samuel Schmidt and Stefan Hilbert},
title = {COSMIC SHEAR RESULTS FROM THE DEEP LENS SURVEY. I. JOINT CONSTRAINTS ON ΩM AND σ8 WITH A TWO-DIMENSIONAL ANALYSIS},
journal = {The Astrophysical Journal}
}

@article{Kim_2019,
doi = {10.3847/1538-4357/ab521e},
url = {https://dx.doi.org/10.3847/1538-4357/ab521e},
year = {2019},
month = {dec},
publisher = {The American Astronomical Society},
volume = {887},
number = {1},
pages = {76},
author = {Jinhyub Kim and M. James Jee and Saul Perlmutter and Brian Hayden and David Rubin and Xiaosheng Huang and Greg Aldering and Jongwan Ko},
title = {Precise Mass Determination of SPT-CL J2106-5844, the Most Massive Cluster at z &gt; 1},
journal = {The Astrophysical Journal}
}

@article{Finner_2023A,
doi = {10.3847/1538-4357/ac9fd3},
url = {https://dx.doi.org/10.3847/1538-4357/ac9fd3},
year = {2022},
month = {dec},
publisher = {The American Astronomical Society},
volume = {942},
number = {1},
pages = {23},
author = {Kyle Finner and Scott W. Randall and M. James Jee and Elizabeth L. Blanton and Hyejeon Cho and Tracy E. Clarke and Simona Giacintucci and Paul Nulsen and Reinout van Weeren},
title = {Hubble Space Telescope and Hyper-Suprime-Cam Weak-lensing Study of the Equal-mass Dissociative Merger CIZA J0107.7+5408},
journal = {The Astrophysical Journal}
}

@article{Finner_2023B,
doi = {10.3847/1538-4357/ace1e6},
url = {https://dx.doi.org/10.3847/1538-4357/ace1e6},
year = {2023},
month = {aug},
publisher = {The American Astronomical Society},
volume = {953},
number = {1},
pages = {102},
author = {Kyle Finner and Andreas Faisst and Ranga-Ram Chary and M. James Jee},
title = {The First Weak-lensing Analysis with the James Webb Space Telescope: SMACS J0723.3–7327},
journal = {The Astrophysical Journal}
}

@article{Mandelbaum_2015,
    author = {Mandelbaum, Rachel and Rowe, Barnaby and Armstrong, Robert and Bard, Deborah and Bertin, Emmanuel and Bosch, James and Boutigny, Dominique and Courbin, Frederic and Dawson, William A. and Donnarumma, Annamaria and Fenech Conti, Ian and Gavazzi, Raphaël and Gentile, Marc and Gill, Mandeep S. S. and Hogg, David W. and Huff, Eric M. and Jee, M. James and Kacprzak, Tomasz and Kilbinger, Martin and Kuntzer, Thibault and Lang, Dustin and Luo, Wentao and March, Marisa C. and Marshall, Philip J. and Meyers, Joshua E. and Miller, Lance and Miyatake, Hironao and Nakajima, Reiko and Ngolé Mboula, Fred Maurice and Nurbaeva, Guldariya and Okura, Yuki and Paulin-Henriksson, Stéphane and Rhodes, Jason and Schneider, Michael D. and Shan, Huanyuan and Sheldon, Erin S. and Simet, Melanie and Starck, Jean-Luc and Sureau, Florent and Tewes, Malte and Zarb Adami, Kristian and Zhang, Jun and Zuntz, Joe},
    title = {GREAT3 results – I. Systematic errors in shear estimation and the impact of real galaxy morphology},
    journal = {Monthly Notices of the Royal Astronomical Society},
    volume = {450},
    number = {3},
    pages = {2963-3007},
    year = {2015},
    month = {05},
    issn = {0035-8711},
    doi = {10.1093/mnras/stv781},
    url = {https://doi.org/10.1093/mnras/stv781},
    eprint = {https://academic.oup.com/mnras/article-pdf/450/3/2963/18508472/stv781.pdf},
}

@ARTICLE{Sevilla-Noarbe_2021,
       author = {{Sevilla-Noarbe}, I. and {Bechtol}, K. and {Carrasco Kind}, M. and {Carnero Rosell}, A. and {Becker}, M.~R. and {Drlica-Wagner}, A. and {Gruendl}, R.~A. and {Rykoff}, E.~S. and {Sheldon}, E. and {Yanny}, B. and {Alarcon}, A. and {Allam}, S. and {Amon}, A. and {Benoit-L{\'e}vy}, A. and {Bernstein}, G.~M. and {Bertin}, E. and {Burke}, D.~L. and {Carretero}, J. and {Choi}, A. and {Diehl}, H.~T. and {Everett}, S. and {Flaugher}, B. and {Gaztanaga}, E. and {Gschwend}, J. and {Harrison}, I. and {Hartley}, W.~G. and {Hoyle}, B. and {Jarvis}, M. and {Johnson}, M.~D. and {Kessler}, R. and {Kron}, R. and {Kuropatkin}, N. and {Leistedt}, B. and {Li}, T.~S. and {Menanteau}, F. and {Morganson}, E. and {Ogando}, R.~L.~C. and {Palmese}, A. and {Paz-Chinch{\'o}n}, F. and {Pieres}, A. and {Pond}, C. and {Rodriguez-Monroy}, M. and {Smith}, J. Allyn and {Stringer}, K.~M. and {Troxel}, M.~A. and {Tucker}, D.~L. and {de Vicente}, J. and {Wester}, W. and {Zhang}, Y. and {Abbott}, T.~M.~C. and {Aguena}, M. and {Annis}, J. and {Avila}, S. and {Bhargava}, S. and {Bridle}, S.~L. and {Brooks}, D. and {Brout}, D. and {Castander}, F.~J. and {Cawthon}, R. and {Chang}, C. and {Conselice}, C. and {Costanzi}, M. and {Crocce}, M. and {da Costa}, L.~N. and {Pereira}, M.~E.~S. and {Davis}, T.~M. and {Desai}, S. and {Dietrich}, J.~P. and {Doel}, P. and {Eckert}, K. and {Evrard}, A.~E. and {Ferrero}, I. and {Fosalba}, P. and {Garc{\'\i}a-Bellido}, J. and {Gerdes}, D.~W. and {Giannantonio}, T. and {Gruen}, D. and {Gutierrez}, G. and {Hinton}, S.~R. and {Hollowood}, D.~L. and {Honscheid}, K. and {Huff}, E.~M. and {Huterer}, D. and {James}, D.~J. and {Jeltema}, T. and {Kuehn}, K. and {Lahav}, O. and {Lidman}, C. and {Lima}, M. and {Lin}, H. and {Maia}, M.~A.~G. and {Marshall}, J.~L. and {Martini}, P. and {Melchior}, P. and {Miquel}, R. and {Mohr}, J.~J. and {Morgan}, R. and {Neilsen}, E. and {Plazas}, A.~A. and {Romer}, A.~K. and {Roodman}, A. and {Sanchez}, E. and {Scarpine}, V. and {Schubnell}, M. and {Serrano}, S. and {Smith}, M. and {Suchyta}, E. and {Tarle}, G. and {Thomas}, D. and {To}, C. and {Varga}, T.~N. and {Wechsler}, R.~H. and {Weller}, J. and {Wilkinson}, R.~D. and {DES Collaboration}},
        title = "{Dark Energy Survey Year 3 Results: Photometric Data Set for Cosmology}",
      journal = {\apjs},
         year = 2021,
        month = jun,
       volume = {254},
       number = {2},
          eid = {24},
        pages = {24},
          doi = {10.3847/1538-4365/abeb66},
archivePrefix = {arXiv},
       eprint = {2011.03407},
 primaryClass = {astro-ph.CO},
       adsurl = {https://ui.adsabs.harvard.edu/abs/2021ApJS..254...24S}
}

@ARTICLE{Clowe_2006,
       author = {{Clowe}, Douglas and {Brada{\v{c}}}, Maru{\v{s}}a and {Gonzalez}, Anthony H. and {Markevitch}, Maxim and {Randall}, Scott W. and {Jones}, Christine and {Zaritsky}, Dennis},
        title = "{A Direct Empirical Proof of the Existence of Dark Matter}",
      journal = {\apjl},
         year = 2006,
        month = sep,
       volume = {648},
       number = {2},
        pages = {L109-L113},
          doi = {10.1086/508162},
archivePrefix = {arXiv},
       eprint = {astro-ph/0608407},
 primaryClass = {astro-ph},
       adsurl = {https://ui.adsabs.harvard.edu/abs/2006ApJ...648L.109C}
}

@ARTICLE{Fischer_1997,
       author = {{Fischer}, Philippe and {Bernstein}, Gary and {Rhee}, George and {Tyson}, J. Anthony},
        title = "{The Mass Distribution of the Cluster 0957+561 From Gravitational Lensing}",
      journal = {\aj},
         year = 1997,
        month = feb,
       volume = {113},
        pages = {521},
          doi = {10.1086/118272},
archivePrefix = {arXiv},
       eprint = {astro-ph/9608117},
 primaryClass = {astro-ph},
       adsurl = {https://ui.adsabs.harvard.edu/abs/1997AJ....113..521F}
}

@article{Wittman_2006,
doi = {10.1086/502621},
url = {https://dx.doi.org/10.1086/502621},
year = {2006},
month = {may},
publisher = {},
volume = {643},
number = {1},
pages = {128},
author = {Wittman, D. and Dell’Antonio, I. P. and Hughes, J. P. and Margoniner, V. E. and Tyson, J. A. and Cohen, J. G. and Norman, D.},
title = {First Results on Shear-selected Clusters from the Deep Lens Survey: Optical Imaging, Spectroscopy, and X-Ray Follow-up},
journal = {The Astrophysical Journal}
}

@ARTICLE{Randall_2008,
       author = {{Randall}, Scott W. and {Markevitch}, Maxim and {Clowe}, Douglas and {Gonzalez}, Anthony H. and {Brada{\v{c}}}, Marusa},
        title = "{Constraints on the Self-Interaction Cross Section of Dark Matter from Numerical Simulations of the Merging Galaxy Cluster 1E 0657-56}",
      journal = {\apj},
         year = 2008,
        month = jun,
       volume = {679},
       number = {2},
        pages = {1173-1180},
          doi = {10.1086/587859},
archivePrefix = {arXiv},
       eprint = {0704.0261},
 primaryClass = {astro-ph},
       adsurl = {https://ui.adsabs.harvard.edu/abs/2008ApJ...679.1173R}
}

@ARTICLE{Navarro_1996,
       author = {{Navarro}, Julio F. and {Frenk}, Carlos S. and {White}, Simon D.~M.},
        title = "{The Structure of Cold Dark Matter Halos}",
      journal = {\apj},
         year = 1996,
        month = may,
       volume = {462},
        pages = {563},
          doi = {10.1086/177173},
archivePrefix = {arXiv},
       eprint = {astro-ph/9508025},
 primaryClass = {astro-ph},
       adsurl = {https://ui.adsabs.harvard.edu/abs/1996ApJ...462..563N}
}

@ARTICLE{Lee_2023,
       author = {{Lee}, Wonki and {Cha}, Sangjun and {Jee}, M. James and {Nagai}, Daisuke and {King}, Lindsay and {ZuHone}, John and {Chadayammuri}, Urmila and {Felix}, Sharon and {Finner}, Kyle},
        title = "{Weak-lensing Mass Bias in Merging Galaxy Clusters}",
      journal = {\apj},
         year = 2023,
        month = mar,
       volume = {945},
       number = {1},
          eid = {71},
        pages = {71},
          doi = {10.3847/1538-4357/acb76b},
archivePrefix = {arXiv},
       eprint = {2211.03892},
 primaryClass = {astro-ph.CO},
       adsurl = {https://ui.adsabs.harvard.edu/abs/2023ApJ...945...71L}
}

@ARTICLE{Foreman-Mackey_2013,
       author = {{Foreman-Mackey}, Daniel and {Hogg}, David W. and {Lang}, Dustin and {Goodman}, Jonathan},
        title = "{emcee: The MCMC Hammer}",
      journal = {\pasp},
         year = 2013,
        month = mar,
       volume = {125},
       number = {925},
        pages = {306},
          doi = {10.1086/670067},
archivePrefix = {arXiv},
       eprint = {1202.3665},
 primaryClass = {astro-ph.IM},
       adsurl = {https://ui.adsabs.harvard.edu/abs/2013PASP..125..306F}
}

@article{Markevitch_2004,
doi = {10.1086/383178},
url = {https://dx.doi.org/10.1086/383178},
year = {2004},
month = {may},
publisher = {},
volume = {606},
number = {2},
pages = {819},
author = {Markevitch, M. and Gonzalez, A. H. and Clowe, D. and Vikhlinin, A. and Forman, W. and Jones, C. and Murray, S. and Tucker, W.},
title = {Direct Constraints on the Dark Matter Self-Interaction Cross Section from the Merging Galaxy Cluster 1E 0657–56},
journal = {The Astrophysical Journal}
}

@article{Clowe_2004,
doi = {10.1086/381970},
url = {https://dx.doi.org/10.1086/381970},
year = {2004},
month = {apr},
publisher = {},
volume = {604},
number = {2},
pages = {596},
author = {Clowe, Douglas and Gonzalez, Anthony and Markevitch, Maxim},
title = {Weak-Lensing Mass Reconstruction of the Interacting Cluster 1E 0657–558: Direct Evidence for the Existence of Dark Matter*},
journal = {The Astrophysical Journal}
}

@INPROCEEDINGS{Sarazin_2002,
       author = {{Sarazin}, Craig L.},
        title = "{The Physics of Cluster Mergers}",
    booktitle = {Merging Processes in Galaxy Clusters},
         year = 2002,
       editor = {{Feretti}, L. and {Gioia}, I.~M. and {Giovannini}, G.},
       series = {Astrophysics and Space Science Library},
       volume = {272},
        month = jun,
        pages = {1-38},
          doi = {10.1007/0-306-48096-4_1},
archivePrefix = {arXiv},
       eprint = {astro-ph/0105418},
 primaryClass = {astro-ph},
       adsurl = {https://ui.adsabs.harvard.edu/abs/2002ASSL..272....1S}
}

@article{Ricker_2001,
doi = {10.1086/323365},
url = {https://dx.doi.org/10.1086/323365},
year = {2001},
month = {nov},
publisher = {},
volume = {561},
number = {2},
pages = {621},
author = {Ricker, Paul M. and Sarazin, Craig L.},
title = {Off-Axis Cluster Mergers: Effects of a Strongly Peaked Dark Matter Profile},
journal = {The Astrophysical Journal}
}

@ARTICLE{Springel_2007,
       author = {{Springel}, Volker and {Farrar}, Glennys R.},
        title = "{The speed of the `bullet' in the merging galaxy cluster 1E0657-56}",
      journal = {\mnras},
         year = 2007,
        month = sep,
       volume = {380},
       number = {3},
        pages = {911-925},
          doi = {10.1111/j.1365-2966.2007.12159.x},
archivePrefix = {arXiv},
       eprint = {astro-ph/0703232},
 primaryClass = {astro-ph},
       adsurl = {https://ui.adsabs.harvard.edu/abs/2007MNRAS.380..911S}
}

@article{Paraficz_2016,
	author = {{Paraficz} and {Kneib} and {Richard} and {Morandi} and {Limousin} and {Jullo} and {Martinez}},
	title = {The Bullet cluster at its best: weighing stars, gas, and dark matter},
	DOI= "10.1051/0004-6361/201527959",
	url= "https://doi.org/10.1051/0004-6361/201527959",
	journal = {A\&A},
	year = 2016,
	volume = 594,
	pages = "A121",
}

@article{Richard_2021,
	author = {{Richard} and {Claeyssens, Adélaïde} and {Lagattuta, David} and {Guaita, Lucia} and {Bauer, Franz Erik} and {Pello, Roser} and {Carton, David} and {Bacon, Roland} and {Soucail, Geneviève} and {Lyon, Gonzalo Prieto} and {Kneib, Jean-Paul} and {Mahler, Guillaume} and {Clément, Benjamin} and {Mercier, Wilfried} and {Variu, Andrei} and {Tamone, Amélie} and {Ebeling, Harald} and {Schmidt, Kasper B.} and {Nanayakkara, Themiya} and {Maseda, Michael} and {Weilbacher, Peter M.} and {Bouché, Nicolas} and {Bouwens, Rychard J.} and {Wisotzki, Lutz} and {de la Vieuville, Geoffroy} and {Martinez, Johany} and {Patrício, Vera}},
	title = {An atlas of MUSE observations towards twelve massive lensing clusters⋆},
	DOI= "10.1051/0004-6361/202039462",
	url= "https://doi.org/10.1051/0004-6361/202039462",
	journal = {A\&A},
	year = 2021,
	volume = 646,
	pages = "A83",
}

@ARTICLE{Melchior_2015,
       author = {{Melchior}, P. and {Suchyta}, E. and {Huff}, E. and {Hirsch}, M. and {Kacprzak}, T. and {Rykoff}, E. and {Gruen}, D. and {Armstrong}, R. and {Bacon}, D. and {Bechtol}, K. and {Bernstein}, G.~M. and {Bridle}, S. and {Clampitt}, J. and {Honscheid}, K. and {Jain}, B. and {Jouvel}, S. and {Krause}, E. and {Lin}, H. and {MacCrann}, N. and {Patton}, K. and {Plazas}, A. and {Rowe}, B. and {Vikram}, V. and {Wilcox}, H. and {Young}, J. and {Zuntz}, J. and {Abbott}, T. and {Abdalla}, F.~B. and {Allam}, S.~S. and {Banerji}, M. and {Bernstein}, J.~P. and {Bernstein}, R.~A. and {Bertin}, E. and {Buckley-Geer}, E. and {Burke}, D.~L. and {Castander}, F.~J. and {da Costa}, L.~N. and {Cunha}, C.~E. and {Depoy}, D.~L. and {Desai}, S. and {Diehl}, H.~T. and {Doel}, P. and {Estrada}, J. and {Evrard}, A.~E. and {Neto}, A. Fausti and {Fernandez}, E. and {Finley}, D.~A. and {Flaugher}, B. and {Frieman}, J.~A. and {Gaztanaga}, E. and {Gerdes}, D. and {Gruendl}, R.~A. and {Gutierrez}, G.~R. and {Jarvis}, M. and {Karliner}, I. and {Kent}, S. and {Kuehn}, K. and {Kuropatkin}, N. and {Lahav}, O. and {Maia}, M.~A.~G. and {Makler}, M. and {Marriner}, J. and {Marshall}, J.~L. and {Merritt}, K.~W. and {Miller}, C.~J. and {Miquel}, R. and {Mohr}, J. and {Neilsen}, E. and {Nichol}, R.~C. and {Nord}, B.~D. and {Reil}, K. and {Roe}, N.~A. and {Roodman}, A. and {Sako}, M. and {Sanchez}, E. and {Santiago}, B.~X. and {Schindler}, R. and {Schubnell}, M. and {Sevilla-Noarbe}, I. and {Sheldon}, E. and {Smith}, C. and {Soares-Santos}, M. and {Swanson}, M.~E.~C. and {Sypniewski}, A.~J. and {Tarle}, G. and {Thaler}, J. and {Thomas}, D. and {Tucker}, D.~L. and {Walker}, A. and {Wechsler}, R. and {Weller}, J. and {Wester}, W.},
        title = "{Mass and galaxy distributions of four massive galaxy clusters from Dark Energy Survey Science Verification data}",
      journal = {\mnras},
         year = 2015,
        month = may,
       volume = {449},
       number = {3},
        pages = {2219-2238},
          doi = {10.1093/mnras/stv398},
archivePrefix = {arXiv},
       eprint = {1405.4285},
 primaryClass = {astro-ph.CO},
       adsurl = {https://ui.adsabs.harvard.edu/abs/2015MNRAS.449.2219M}
}

@article{Milosavljevic_2007,
doi = {10.1086/518960},
url = {https://dx.doi.org/10.1086/518960},
year = {2007},
month = {may},
publisher = {},
volume = {661},
number = {2},
pages = {L131},
author = {Milosavljević, Miloš and Koda, Jun and Nagai, Daisuke and Nakar, Ehud and Shapiro, Paul R.},
title = {The Cluster-Merger Shock in 1E 0657–56: Faster than a Speeding Bullet?},
journal = {The Astrophysical Journal}
}

@article{Mastropietro_2008,
    author = {Mastropietro, Chiara and Burkert, Andreas},
    title = {Simulating the Bullet Cluster},
    journal = {Monthly Notices of the Royal Astronomical Society},
    volume = {389},
    number = {2},
    pages = {967-988},
    year = {2008},
    month = {09},
    issn = {0035-8711},
    doi = {10.1111/j.1365-2966.2008.13626.x},
    url = {https://doi.org/10.1111/j.1365-2966.2008.13626.x},
    eprint = {https://academic.oup.com/mnras/article-pdf/389/2/967/2970173/mnras0389-0967.pdf},
}

@article{Lage_2014,
doi = {10.1088/0004-637X/787/2/144},
url = {https://dx.doi.org/10.1088/0004-637X/787/2/144},
year = {2014},
month = {may},
publisher = {The American Astronomical Society},
volume = {787},
number = {2},
pages = {144},
author = {Lage, Craig and Farrar, Glennys},
title = {CONSTRAINED SIMULATION OF THE BULLET CLUSTER},
journal = {The Astrophysical Journal}
}

@ARTICLE{Wu_2019,
       author = {{Wu}, Hao-Yi and {Weinberg}, David H. and {Salcedo}, Andr{\'e}s N. and {Wibking}, Benjamin D. and {Zu}, Ying},
        title = "{Covariance matrices for galaxy cluster weak lensing: from virial regime to uncorrelated large-scale structure}",
      journal = {\mnras},
         year = 2019,
        month = dec,
       volume = {490},
       number = {2},
        pages = {2606-2626},
          doi = {10.1093/mnras/stz2617},
archivePrefix = {arXiv},
       eprint = {1907.06611},
 primaryClass = {astro-ph.CO},
       adsurl = {https://ui.adsabs.harvard.edu/abs/2019MNRAS.490.2606W}
}

@ARTICLE{Diemer_2019,
       author = {{Diemer}, Benedikt and {Joyce}, Michael},
        title = "{An Accurate Physical Model for Halo Concentrations}",
      journal = {\apj},
         year = 2019,
        month = feb,
       volume = {871},
       number = {2},
          eid = {168},
        pages = {168},
          doi = {10.3847/1538-4357/aafad6},
archivePrefix = {arXiv},
       eprint = {1809.07326},
 primaryClass = {astro-ph.CO},
       adsurl = {https://ui.adsabs.harvard.edu/abs/2019ApJ...871..168D}
}

@article{Oguri_2011,
    author = {Oguri, Masamune and Hamana, Takashi},
    title = {Detailed cluster lensing profiles at large radii and the impact on cluster weak lensing studies},
    journal = {Monthly Notices of the Royal Astronomical Society},
    volume = {414},
    number = {3},
    pages = {1851-1861},
    year = {2011},
    month = {06},
    issn = {0035-8711},
    doi = {10.1111/j.1365-2966.2011.18481.x},
    url = {https://doi.org/10.1111/j.1365-2966.2011.18481.x},
    eprint = {https://academic.oup.com/mnras/article-pdf/414/3/1851/3472144/mnras0414-1851.pdf},
}

@article{Jee_2014,
doi = {10.1088/0004-637X/785/1/20},
url = {https://dx.doi.org/10.1088/0004-637X/785/1/20},
year = {2014},
month = {mar},
publisher = {The American Astronomical Society},
volume = {785},
number = {1},
pages = {20},
author = {Jee, M. James and Hughes, John P. and Menanteau, Felipe and Sifón, Cristóbal and Mandelbaum, Rachel and Barrientos, L. Felipe and Infante, Leopoldo and Ng, Karen Y.},
title = {WEIGHING “EL GORDO” WITH A PRECISION SCALE: HUBBLE SPACE TELESCOPE WEAK-LENSING ANALYSIS OF THE MERGING GALAXY CLUSTER ACT-CL J0102−4915 AT z = 0.87},
journal = {The Astrophysical Journal}
}

@ARTICLE{Sikhosana_2023,
       author = {{Sikhosana}, S.~P. and {Knowles}, K. and {Hilton}, M. and {Moodley}, K. and {Murgia}, M.},
        title = "{MeerKAT's view of the bullet cluster 1E 0657-55.8}",
      journal = {\mnras},
         year = 2023,
        month = jan,
       volume = {518},
       number = {3},
        pages = {4595-4605},
          doi = {10.1093/mnras/stac3370},
archivePrefix = {arXiv},
       eprint = {2207.05492},
 primaryClass = {astro-ph.GA},
       adsurl = {https://ui.adsabs.harvard.edu/abs/2023MNRAS.518.4595S}
}

@ARTICLE{Finner_2025,
       author = {{Finner}, Kyle and {Jee}, M. James and {Cho}, Hyejeon and {HyeongHan}, Kim and {Lee}, Wonki and {van Weeren}, Reinout J. and {Wittman}, David and {Yoon}, Mijin},
        title = "{Weak-lensing Characterization of the Dark Matter in 29 Merging Clusters that Exhibit Radio Relics}",
      journal = {\apjs},
         year = 2025,
        month = mar,
       volume = {277},
       number = {1},
          eid = {28},
        pages = {28},
          doi = {10.3847/1538-4365/adb0b6},
archivePrefix = {arXiv},
       eprint = {2407.02557},
 primaryClass = {astro-ph.CO},
       adsurl = {https://ui.adsabs.harvard.edu/abs/2025ApJS..277...28F}
}

@ARTICLE{Osato_2021,
       author = {{Osato}, Ken and {Liu}, Jia and {Haiman}, Zolt{\'a}n},
        title = "{{\ensuremath{\kappa}}TNG: effect of baryonic processes on weak lensing with IllustrisTNG simulations}",
      journal = {\mnras},
         year = 2021,
        month = apr,
       volume = {502},
       number = {4},
        pages = {5593-5602},
          doi = {10.1093/mnras/stab395},
archivePrefix = {arXiv},
       eprint = {2010.09731},
 primaryClass = {astro-ph.CO},
       adsurl = {https://ui.adsabs.harvard.edu/abs/2021MNRAS.502.5593O}
}

@INPROCEEDINGS{Markevitch_2006,
       author = {{Markevitch}, M.},
        title = "{Chandra Observation of the Most Interesting Cluster in the Universe}",
    booktitle = {The X-ray Universe 2005},
         year = 2006,
       editor = {{Wilson}, A.},
       series = {ESA Special Publication},
       volume = {604},
        month = jan,
        pages = {723},
          doi = {10.48550/arXiv.astro-ph/0511345},
archivePrefix = {arXiv},
       eprint = {astro-ph/0511345},
 primaryClass = {astro-ph},
       adsurl = {https://ui.adsabs.harvard.edu/abs/2006ESASP.604..723M}
}

@ARTICLE{Hilton_2021,
       author = {{Hilton}, M. and {Sif{\'o}n}, C. and {Naess}, S. and {Madhavacheril}, M. and {Oguri}, M. and {Rozo}, E. and {Rykoff}, E. and {Abbott}, T.~M.~C. and {Adhikari}, S. and {Aguena}, M. and {Aiola}, S. and {Allam}, S. and {Amodeo}, S. and {Amon}, A. and {Annis}, J. and {Ansarinejad}, B. and {Aros-Bunster}, C. and {Austermann}, J.~E. and {Avila}, S. and {Bacon}, D. and {Battaglia}, N. and {Beall}, J.~A. and {Becker}, D.~T. and {Bernstein}, G.~M. and {Bertin}, E. and {Bhandarkar}, T. and {Bhargava}, S. and {Bond}, J.~R. and {Brooks}, D. and {Burke}, D.~L. and {Calabrese}, E. and {Carrasco Kind}, M. and {Carretero}, J. and {Choi}, S.~K. and {Choi}, A. and {Conselice}, C. and {da Costa}, L.~N. and {Costanzi}, M. and {Crichton}, D. and {Crowley}, K.~T. and {D{\"u}nner}, R. and {Denison}, E.~V. and {Devlin}, M.~J. and {Dicker}, S.~R. and {Diehl}, H.~T. and {Dietrich}, J.~P. and {Doel}, P. and {Duff}, S.~M. and {Duivenvoorden}, A.~J. and {Dunkley}, J. and {Everett}, S. and {Ferraro}, S. and {Ferrero}, I. and {Fert{\'e}}, A. and {Flaugher}, B. and {Frieman}, J. and {Gallardo}, P.~A. and {Garc{\'\i}a-Bellido}, J. and {Gaztanaga}, E. and {Gerdes}, D.~W. and {Giles}, P. and {Golec}, J.~E. and {Gralla}, M.~B. and {Grandis}, S. and {Gruen}, D. and {Gruendl}, R.~A. and {Gschwend}, J. and {Gutierrez}, G. and {Han}, D. and {Hartley}, W.~G. and {Hasselfield}, M. and {Hill}, J.~C. and {Hilton}, G.~C. and {Hincks}, A.~D. and {Hinton}, S.~R. and {Ho}, S. -P.~P. and {Honscheid}, K. and {Hoyle}, B. and {Hubmayr}, J. and {Huffenberger}, K.~M. and {Hughes}, J.~P. and {Jaelani}, A.~T. and {Jain}, B. and {James}, D.~J. and {Jeltema}, T. and {Kent}, S. and {Knowles}, K. and {Koopman}, B.~J. and {Kuehn}, K. and {Lahav}, O. and {Lima}, M. and {Lin}, Y. -T. and {Lokken}, M. and {Loubser}, S.~I. and {MacCrann}, N. and {Maia}, M.~A.~G. and {Marriage}, T.~A. and {Martin}, J. and {McMahon}, J. and {Melchior}, P. and {Menanteau}, F. and {Miquel}, R. and {Miyatake}, H. and {Moodley}, K. and {Morgan}, R. and {Mroczkowski}, T. and {Nati}, F. and {Newburgh}, L.~B. and {Niemack}, M.~D. and {Nishizawa}, A.~J. and {Ogando}, R.~L.~C. and {Orlowski-Scherer}, J. and {Page}, L.~A. and {Palmese}, A. and {Partridge}, B. and {Paz-Chinch{\'o}n}, F. and {Phakathi}, P. and {Plazas}, A.~A. and {Robertson}, N.~C. and {Romer}, A.~K. and {Carnero Rosell}, A. and {Salatino}, M. and {Sanchez}, E. and {Schaan}, E. and {Schillaci}, A. and {Sehgal}, N. and {Serrano}, S. and {Shin}, T. and {Simon}, S.~M. and {Smith}, M. and {Soares-Santos}, M. and {Spergel}, D.~N. and {Staggs}, S.~T. and {Storer}, E.~R. and {Suchyta}, E. and {Swanson}, M.~E.~C. and {Tarle}, G. and {Thomas}, D. and {To}, C. and {Trac}, H. and {Ullom}, J.~N. and {Vale}, L.~R. and {Van Lanen}, J. and {Vavagiakis}, E.~M. and {De Vicente}, J. and {Wilkinson}, R.~D. and {Wollack}, E.~J. and {Xu}, Z. and {Zhang}, Y.},
        title = "{The Atacama Cosmology Telescope: A Catalog of >4000 Sunyaev-Zel{\textquoteright}dovich Galaxy Clusters}",
      journal = {\apjs},
         year = 2021,
        month = mar,
       volume = {253},
       number = {1},
          eid = {3},
        pages = {3},
          doi = {10.3847/1538-4365/abd023},
archivePrefix = {arXiv},
       eprint = {2009.11043},
 primaryClass = {astro-ph.CO},
       adsurl = {https://ui.adsabs.harvard.edu/abs/2021ApJS..253....3H}
}

@ARTICLE{Walker_2025,
       author = {{Walker}, Kris and {Ludlow}, Aaron and {Power}, Chris and {Knebe}, Alexander and {Cui}, Weiguang},
        title = "{The Three Hundred Project: deducing the stellar splashback structure of galaxy clusters from their orbiting profiles}",
      journal = {arXiv e-prints},
         year = 2025,
        month = aug,
          eid = {arXiv:2508.07232},
        pages = {arXiv:2508.07232},
          doi = {10.48550/arXiv.2508.07232},
archivePrefix = {arXiv},
       eprint = {2508.07232},
 primaryClass = {astro-ph.GA},
       adsurl = {https://ui.adsabs.harvard.edu/abs/2025arXiv250807232W}
}

@ARTICLE{Kaiser_1993,
       author = {{Kaiser}, Nick and {Squires}, Gordon},
        title = "{Mapping the Dark Matter with Weak Gravitational Lensing}",
      journal = {\apj},
         year = 1993,
        month = feb,
       volume = {404},
        pages = {441},
          doi = {10.1086/172297},
       adsurl = {https://ui.adsabs.harvard.edu/abs/1993ApJ...404..441K}
}

@ARTICLE{Rowe_2010,
       author = {{Rowe}, Barnaby},
        title = "{Improving PSF modelling for weak gravitational lensing using new methods in model selection}",
      journal = {\mnras},
         year = 2010,
        month = may,
       volume = {404},
       number = {1},
        pages = {350-366},
          doi = {10.1111/j.1365-2966.2010.16277.x},
archivePrefix = {arXiv},
       eprint = {0904.3056},
 primaryClass = {astro-ph.CO},
       adsurl = {https://ui.adsabs.harvard.edu/abs/2010MNRAS.404..350R}
}

@ARTICLE{Eckert_2022,
       author = {{Eckert}, D. and {Ettori}, S. and {Robertson}, A. and {Massey}, R. and {Pointecouteau}, E. and {Harvey}, D. and {McCarthy}, I.~G.},
        title = "{Constraints on dark matter self-interaction from the internal density profiles of X-COP galaxy clusters}",
      journal = {\aap},
         year = 2022,
        month = oct,
       volume = {666},
          eid = {A41},
        pages = {A41},
          doi = {10.1051/0004-6361/202243205},
archivePrefix = {arXiv},
       eprint = {2205.01123},
 primaryClass = {astro-ph.CO},
       adsurl = {https://ui.adsabs.harvard.edu/abs/2022A&A...666A..41E}
}

@software{Bushouse_2024,
    author = {Bushouse, Howard and Eisenhamer, Jonathan and Dencheva, Nadia and Davies, James and Greenfield, Perry and Morrison, Jane and Hodge, Phil and Simon, Bernie and Grumm, David and Droettboom, Michael and Slavich, Edward and Sosey, Megan and Pauly, Tyler and Miller, Todd and Jedrzejewski, Robert and Hack, Warren and Davis, David and Crawford, Steven and Law, David and Gordon, Karl and Regan, Michael and Cara, Mihai and MacDonald, Ken and Bradley, Larry and Shanahan, Clare and Jamieson, William and Teodoro, Mairan and Williams, Thomas and Pena-Guerrero, Maria and Graham, Brett and Molter, Edward and Brandt, Timothy and Hayes, Christian and Cooper, Rachel and Clarke, Melanie},
    doi = {10.5281/zenodo.7038885},
    month = sep,
    title = {{JWST Calibration Pipeline}},
    url = {https://github.com/spacetelescope/jwst},
    version = {1.16.0},
    year = {2024}
}

@ARTICLE{Bagley_2023,
       author = {{Bagley}, Micaela B. and {Finkelstein}, Steven L. and {Koekemoer}, Anton M. and {Ferguson}, Henry C. and {Arrabal Haro}, Pablo and {Dickinson}, Mark and {Kartaltepe}, Jeyhan S. and {Papovich}, Casey and {P{\'e}rez-Gonz{\'a}lez}, Pablo G. and {Pirzkal}, Nor and {Somerville}, Rachel S. and {Willmer}, Christopher N.~A. and {Yang}, Guang and {Yung}, L.~Y. Aaron and {Fontana}, Adriano and {Grazian}, Andrea and {Grogin}, Norman A. and {Hirschmann}, Michaela and {Kewley}, Lisa J. and {Kirkpatrick}, Allison and {Kocevski}, Dale D. and {Lotz}, Jennifer M. and {Medrano}, Aubrey and {Morales}, Alexa M. and {Pentericci}, Laura and {Ravindranath}, Swara and {Trump}, Jonathan R. and {Wilkins}, Stephen M. and {Calabr{\`o}}, Antonello and {Cooper}, M.~C. and {Costantin}, Luca and {de la Vega}, Alexander and {Hilbert}, Bryan and {Hutchison}, Taylor A. and {Larson}, Rebecca L. and {Lucas}, Ray A. and {McGrath}, Elizabeth J. and {Ryan}, Russell and {Wang}, Xin and {Wuyts}, Stijn},
        title = "{CEERS Epoch 1 NIRCam Imaging: Reduction Methods and Simulations Enabling Early JWST Science Results}",
      journal = {\apjl},
         year = 2023,
        month = mar,
       volume = {946},
       number = {1},
          eid = {L12},
        pages = {L12},
          doi = {10.3847/2041-8213/acbb08},
archivePrefix = {arXiv},
       eprint = {2211.02495},
 primaryClass = {astro-ph.IM},
       adsurl = {https://ui.adsabs.harvard.edu/abs/2023ApJ...946L..12B}
}

@ARTICLE{Brammer_2008,
       author = {{Brammer}, Gabriel B. and {van Dokkum}, Pieter G. and {Coppi}, Paolo},
        title = "{EAZY: A Fast, Public Photometric Redshift Code}",
      journal = {\apj},
         year = 2008,
        month = oct,
       volume = {686},
       number = {2},
        pages = {1503-1513},
          doi = {10.1086/591786},
archivePrefix = {arXiv},
       eprint = {0807.1533},
 primaryClass = {astro-ph},
       adsurl = {https://ui.adsabs.harvard.edu/abs/2008ApJ...686.1503B}
}

@ARTICLE{Golovich_2019,
       author = {{Golovich}, N. and {Dawson}, W.~A. and {Wittman}, D.~M. and {van Weeren}, R.~J. and {Andrade-Santos}, F. and {Jee}, M.~J. and {Benson}, B. and {de Gasperin}, F. and {Venturi}, T. and {Bonafede}, A. and {Sobral}, D. and {Ogrean}, G.~A. and {Lemaux}, B.~C. and {Brada{\v{c}}}, M. and {Br{\"u}ggen}, M. and {Peter}, A.},
        title = "{Merging Cluster Collaboration: A Panchromatic Atlas of Radio Relic Mergers}",
      journal = {\apj},
         year = 2019,
        month = sep,
       volume = {882},
       number = {1},
          eid = {69},
        pages = {69},
          doi = {10.3847/1538-4357/ab2f90},
archivePrefix = {arXiv},
       eprint = {1806.10619},
 primaryClass = {astro-ph.CO},
       adsurl = {https://ui.adsabs.harvard.edu/abs/2019ApJ...882...69G}
}

@ARTICLE{Cha_2022,
       author = {{Cha}, Sangjun and {Jee}, M. James},
        title = "{MARS: A New Maximum-entropy-regularized Strong Lensing Mass Reconstruction Method}",
      journal = {\apj},
         year = 2022,
        month = jun,
       volume = {931},
       number = {2},
          eid = {127},
        pages = {127},
          doi = {10.3847/1538-4357/ac69df},
archivePrefix = {arXiv},
       eprint = {2202.10489},
 primaryClass = {astro-ph.CO},
       adsurl = {https://ui.adsabs.harvard.edu/abs/2022ApJ...931..127C}
}

@article{Gronau_2017,
title = {A tutorial on bridge sampling},
journal = {Journal of Mathematical Psychology},
volume = {81},
pages = {80-97},
year = {2017},
issn = {0022-2496},
doi = {https://doi.org/10.1016/j.jmp.2017.09.005},
url = {https://www.sciencedirect.com/science/article/pii/S0022249617300640},
author = {Quentin F. Gronau and Alexandra Sarafoglou and Dora Matzke and Alexander Ly and Udo Boehm and Maarten Marsman and David S. Leslie and Jonathan J. Forster and Eric-Jan Wagenmakers and Helen Steingroever}
}

@article{Newton_1994,
    author = {Newton, Michael A. and Raftery, Adrian E.},
    title = {Approximate Bayesian Inference with the Weighted Likelihood Bootstrap},
    journal = {Journal of the Royal Statistical Society: Series B (Methodological)},
    volume = {56},
    number = {1},
    pages = {3-26},
    year = {2018},
    month = {12},
    issn = {0035-9246},
    doi = {10.1111/j.2517-6161.1994.tb01956.x},
    url = {https://doi.org/10.1111/j.2517-6161.1994.tb01956.x},
    eprint = {https://academic.oup.com/jrsssb/article-pdf/56/1/3/49173498/jrsssb_56_1_3.pdf},
}

@INPROCEEDINGS{Robert_2009,
       author = {{Robert}, C.~P. and {Wraith}, D.},
        title = "{Computational methods for Bayesian model choice}",
    booktitle = {Bayesian Inference and Maximum Entropy Methods in Science and Engineering: The 29th International Workshop on Bayesian Inference and Maximum Entropy Methods in Science and Engineering},
         year = 2009,
       editor = {{Goggans}, Paul M. and {Chan}, Chun-Yong},
       series = {American Institute of Physics Conference Series},
       volume = {1193},
        month = dec,
    publisher = {AIP},
        pages = {251-262},
          doi = {10.1063/1.3275622},
archivePrefix = {arXiv},
       eprint = {0907.5123},
 primaryClass = {stat.CO},
       adsurl = {https://ui.adsabs.harvard.edu/abs/2009AIPC.1193..251R}
}

@article{Gelfand_1994,
 ISSN = {00359246},
 URL = {http://www.jstor.org/stable/2346123},
 author = {A. E. Gelfand and D. K. Dey},
 journal = {Journal of the Royal Statistical Society. Series B (Methodological)},
 number = {3},
 pages = {501--514},
 publisher = {[Royal Statistical Society, Oxford University Press]},
 title = {Bayesian Model Choice: Asymptotics and Exact Calculations},
 urldate = {2025-11-03},
 volume = {56},
 year = {1994}
}

@ARTICLE{Lee_2025,
       author = {{Lee}, Wonki and {Pillepich}, Annalisa and {Nelson}, Dylan and {Jee}, Myungkook James and {Nagai}, Daisuke and {Finner}, Kyle and {ZuHone}, John},
        title = "{Exploring the statistical properties of double radio relics in the TNG-Cluster and TNG300 simulations}",
      journal = {arXiv e-prints},
         year = 2025,
        month = oct,
          eid = {arXiv:2510.21632},
        pages = {arXiv:2510.21632},
          doi = {10.48550/arXiv.2510.21632},
archivePrefix = {arXiv},
       eprint = {2510.21632},
 primaryClass = {astro-ph.GA},
       adsurl = {https://ui.adsabs.harvard.edu/abs/2025arXiv251021632L}
}

@ARTICLE{Knowles_2022,
       author = {{Knowles}, K. and {Cotton}, W.~D. and {Rudnick}, L. and {Camilo}, F. and {Goedhart}, S. and {Deane}, R. and {Ramatsoku}, M. and {Bietenholz}, M.~F. and {Br{\"u}ggen}, M. and {Button}, C. and {Chen}, H. and {Chibueze}, J.~O. and {Clarke}, T.~E. and {de Gasperin}, F. and {Ianjamasimanana}, R. and {J{\'o}zsa}, G.~I.~G. and {Hilton}, M. and {Kesebonye}, K.~C. and {Kolokythas}, K. and {Kraan-Korteweg}, R.~C. and {Lawrie}, G. and {Lochner}, M. and {Loubser}, S.~I. and {Marchegiani}, P. and {Mhlahlo}, N. and {Moodley}, K. and {Murphy}, E. and {Namumba}, B. and {Oozeer}, N. and {Parekh}, V. and {Pillay}, D.~S. and {Passmoor}, S.~S. and {Ramaila}, A.~J.~T. and {Ranchod}, S. and {Retana-Montenegro}, E. and {Sebokolodi}, L. and {Sikhosana}, S.~P. and {Smirnov}, O. and {Thorat}, K. and {Venturi}, T. and {Abbott}, T.~D. and {Adam}, R.~M. and {Adams}, G. and {Aldera}, M.~A. and {Bauermeister}, E.~F. and {Bennett}, T.~G.~H. and {Bode}, W.~A. and {Botha}, D.~H. and {Botha}, A.~G. and {Brederode}, L.~R.~S. and {Buchner}, S. and {Burger}, J.~P. and {Cheetham}, T. and {de Villiers}, D.~I.~L. and {Dikgale-Mahlakoana}, M.~A. and {du Toit}, L.~J. and {Esterhuyse}, S.~W.~P. and {Fadana}, G. and {Fanaroff}, B.~L. and {Fataar}, S. and {Foley}, A.~R. and {Fourie}, D.~J. and {Frank}, B.~S. and {Gamatham}, R.~R.~G. and {Gatsi}, T.~G. and {Geyer}, M. and {Gouws}, M. and {Gumede}, S.~C. and {Heywood}, I. and {Hlakola}, M.~J. and {Hokwana}, A. and {Hoosen}, S.~W. and {Horn}, D.~M. and {Horrell}, J.~M.~G. and {Hugo}, B.~V. and {Isaacson}, A.~R. and {Jonas}, J.~L. and {Jordaan}, J.~D.~B. and {Joubert}, A.~F. and {Julie}, R.~P.~M. and {Kapp}, F.~B. and {Kasper}, V.~A. and {Kenyon}, J.~S. and {Kotz{\'e}}, P.~P.~A. and {Kotze}, A.~G. and {Kriek}, N. and {Kriel}, H. and {Krishnan}, V.~K. and {Kusel}, T.~W. and {Legodi}, L.~S. and {Lehmensiek}, R. and {Liebenberg}, D. and {Lord}, R.~T. and {Lunsky}, B.~M. and {Madisa}, K. and {Magnus}, L.~G. and {Main}, J.~P.~L. and {Makhaba}, A. and {Makhathini}, S. and {Malan}, J.~A. and {Manley}, J.~R. and {Marais}, S.~J. and {Maree}, M.~D.~J. and {Martens}, A. and {Mauch}, T. and {McAlpine}, K. and {Merry}, B.~C. and {Millenaar}, R.~P. and {Mokone}, O.~J. and {Monama}, T.~E. and {Mphego}, M.~C. and {New}, W.~S. and {Ngcebetsha}, B. and {Ngoasheng}, K.~J. and {Ockards}, M.~T. and {Otto}, A.~J. and {Patel}, A.~A. and {Peens-Hough}, A. and {Perkins}, S.~J. and {Ramanujam}, N.~M. and {Ramudzuli}, Z.~R. and {Ratcliffe}, S.~M. and {Renil}, R. and {Robyntjies}, A. and {Rust}, A.~N. and {Salie}, S. and {Sambu}, N. and {Schollar}, C.~T.~G. and {Schwardt}, L.~C. and {Schwartz}, R.~L. and {Serylak}, M. and {Siebrits}, R. and {Sirothia}, S.~K. and {Slabber}, M. and {Sofeya}, L. and {Taljaard}, B. and {Tasse}, C. and {Tiplady}, A.~J. and {Toruvanda}, O. and {Twum}, S.~N. and {van Balla}, T.~J. and {van der Byl}, A. and {van der Merwe}, C. and {van Dyk}, C.~L. and {Van Tonder}, V. and {Van Wyk}, R. and {Venter}, A.~J. and {Venter}, M. and {Welz}, M.~G. and {Williams}, L.~P. and {Xaia}, B.},
        title = "{The MeerKAT Galaxy Cluster Legacy Survey. I. Survey Overview and Highlights}",
      journal = {\aap},
         year = 2022,
        month = jan,
       volume = {657},
          eid = {A56},
        pages = {A56},
          doi = {10.1051/0004-6361/202141488},
archivePrefix = {arXiv},
       eprint = {2111.05673},
 primaryClass = {astro-ph.GA},
       adsurl = {https://ui.adsabs.harvard.edu/abs/2022A&A...657A..56K}
}

@ARTICLE{Einasto_1965,
       author = {{Einasto}, J.},
        title = "{On the Construction of a Composite Model for the Galaxy and on the Determination of the System of Galactic Parameters}",
      journal = {Trudy Astrofizicheskogo Instituta Alma-Ata},
         year = 1965,
        month = jan,
       volume = {5},
        pages = {87-100},
       adsurl = {https://ui.adsabs.harvard.edu/abs/1965TrAlm...5...87E}
}

@ARTICLE{Seitz_1997,
       author = {{Seitz}, C. and {Schneider}, P.},
        title = "{Steps towards nonlinear cluster inversion through gravitational distortions. III. Including a redshift distribution of the sources.}",
      journal = {\aap},
         year = 1997,
        month = feb,
       volume = {318},
        pages = {687-699},
          doi = {10.48550/arXiv.astro-ph/9601079},
archivePrefix = {arXiv},
       eprint = {astro-ph/9601079},
 primaryClass = {astro-ph},
       adsurl = {https://ui.adsabs.harvard.edu/abs/1997A&A...318..687S}
}

@ARTICLE{Robertson_2017,
       author = {{Robertson}, Andrew and {Massey}, Richard and {Eke}, Vincent},
        title = "{What does the Bullet Cluster tell us about self-interacting dark matter?}",
      journal = {\mnras},
         year = 2017,
        month = feb,
       volume = {465},
       number = {1},
        pages = {569-587},
          doi = {10.1093/mnras/stw2670},
archivePrefix = {arXiv},
       eprint = {1605.04307},
 primaryClass = {astro-ph.CO},
       adsurl = {https://ui.adsabs.harvard.edu/abs/2017MNRAS.465..569R}
}

@article{Vecchi_2025,
	author = {{Vecchi} and {Harvey} and {Nightingale} and {Schaller} and {Schaye} and {Tregidga}},
	title = {Impact of line of sight structure on weak lensing observables of galaxy clusters},
	DOI= "10.1051/0004-6361/202555162",
	url= "https://doi.org/10.1051/0004-6361/202555162",
	journal = {A\&A},
	year = 2025,
	volume = 703,
	pages = "A45",
}

@article{Benavides_2023,
	author = {{Benavides} and {Biviano} and {Abadi}},
	title = {DS+: A method for the identification of cluster substructures},
	DOI= "10.1051/0004-6361/202245422",
	url= "https://doi.org/10.1051/0004-6361/202245422",
	journal = {A\&A},
	year = 2023,
	volume = 669,
	pages = "A147",
}

@article{Harvey_2015,
author = {David Harvey  and Richard Massey  and Thomas Kitching  and Andy Taylor  and Eric Tittley },
title = {The nongravitational interactions of dark matter in colliding galaxy clusters},
journal = {Science},
volume = {347},
number = {6229},
pages = {1462-1465},
year = {2015},
doi = {10.1126/science.1261381},
URL = {https://www.science.org/doi/abs/10.1126/science.1261381},
eprint = {https://www.science.org/doi/pdf/10.1126/science.1261381}}

@article{Kim_2017,
    author = {Kim, Stacy Y. and Peter, Annika H. G. and Wittman, David},
    title = {In the wake of dark giants: new signatures of dark matter self-interactions in equal-mass mergers of galaxy clusters},
    journal = {Monthly Notices of the Royal Astronomical Society},
    volume = {469},
    number = {2},
    pages = {1414-1444},
    year = {2017},
    month = {08},
    issn = {0035-8711},
    doi = {10.1093/mnras/stx896},
    url = {https://doi.org/10.1093/mnras/stx896},
    eprint = {https://academic.oup.com/mnras/article-pdf/469/2/1414/17272506/stx896.pdf},
}

@ARTICLE{Jee_2026,
       author = {{Jee}, M. James and {Park}, Dongak and {Lee}, Wonki},
        title = "{A New Robust Constraint on the Self-interaction Cross-section of Dark Matter with Double Radio Relic Clusters}",
      journal = {arXiv e-prints},
         year = 2026,
        month = apr,
          eid = {arXiv:2605.00093},
        pages = {arXiv:2605.00093},
          doi = {10.48550/arXiv.2605.00093},
archivePrefix = {arXiv},
       eprint = {2605.00093},
 primaryClass = {astro-ph.CO},
       adsurl = {https://ui.adsabs.harvard.edu/abs/2026arXiv260500093J}
}

\end{document}